\documentclass{article}
\usepackage[backend=bibtex]{biblatex}
\usepackage[margin=1in]{geometry}
\usepackage{appendix}
\usepackage{authblk}
\usepackage{booktabs}

\usepackage{graphicx}
\usepackage{soul}
\usepackage[dvipsnames]{xcolor}
\usepackage{enumitem}
\usepackage{array}
\usepackage{scrextend}
\usepackage{caption}
\usepackage{booktabs}
\usepackage{makecell}
\usepackage{nicematrix}
\usepackage{tikz}
\usetikzlibrary{shapes.geometric, arrows}
\usepackage{hyperref}
\usepackage{tabularx}
\usetikzlibrary{calc}

\newif\ifcomments\commentsfalse
\newif\ifanon\anontrue
\newif\iffullversion\fullversiontrue

\ifcomments
    \newcommand{\andres}[1]{{\textcolor{BrickRed}{AF: #1}}}
    \newcommand{\betty}[1]{{\textcolor{TealBlue}{BLH: #1}}}
    \newcommand{\daniella}[1]{{\textcolor{Periwinkle}{DF: #1}}}
    \newcommand{\derek}[1]{{\textcolor{BurntOrange}{DY: #1}}}
    \newcommand{\jacob}[1]{{\textcolor{WildStrawberry}{JL: #1}}}
    \newcommand{\kyunghyun}[1]{{\textcolor{Mahogany}{KC: #1}}}
    \newcommand{\mallory}[1]{{\textcolor{ForestGreen}{MK: #1}}}
    \newcommand{\sam}[1]{{\textcolor{Plum}{SA: #1}}}
    \newcommand{\sunoo}[1]{{\textcolor{NavyBlue}{SP: #1}}}
\else
    \newcommand{\andres}[1]{\ignorespaces}
    \newcommand{\betty}[1]{\ignorespaces}
    \newcommand{\daniella}[1]{\ignorespaces}
    \newcommand{\derek}[1]{\ignorespaces}
    \newcommand{\jacob}[1]{\ignorespaces}
    \newcommand{\kyunghyun}[1]{\ignorespaces}
    \newcommand{\mallory}[1]{\ignorespaces}
    \newcommand{\sam}[1]{\ignorespaces}
    \newcommand{\sunoo}[1]{\ignorespaces}
\fi

\newcommand{\bi}[1]{\textbf{\textit{#1}}}
\newcommand{\subh}[1]{\smallskip \noindent \textbf{{#1}}.}
\renewcommand{\paragraph}[1]{\subh{#1}}

\newtheorem{recommendation}{Recommendation}

\newlength{\saveparindent}
\newlength{\saveparskip}
\newcounter{ctr}

\newenvironment{newitemize}{%
\begin{list}{\mbox{}\hspace{5pt}$\bullet$\hfill}{\labelwidth=15pt%
\labelsep=5pt \leftmargin=20pt \topsep=3pt%
\setlength{\listparindent}{\saveparindent}%
\setlength{\parsep}{\saveparskip}%
\setlength{\itemsep}{3pt} }}{\end{list}}

\newcommand{\nocontentsline}[3]{}
\let\origcontentsline\addcontentsline
\newcommand\stoptoc{\let\addcontentsline\nocontentsline}
\newcommand\resumetoc{\let\addcontentsline\origcontentsline}

\title{How To Think About End-To-End Encryption and AI: \\ Training, Processing, Disclosure, and Consent}
\author[1]{Mallory Knodel}
\author[2]{Andr\'{e}s F\'{a}brega}
\author[1]{Daniella Ferrari}
\author[1]{Jacob Leiken}
\author[1]{Betty Li Hou}
\author[1]{\\Derek Yen}
\author[1]{Sam de Alfaro}
\author[1]{Kyunghyun Cho}
\author[1]{Sunoo Park}
\affil[1]{New York University}
\affil[2]{Cornell University}
\date{March 2025}

\begin{document}

\maketitle

\begin{abstract}
End-to-end encryption (E2EE) has become the gold standard for securing communications, bringing strong confidentiality and privacy guarantees to billions of users worldwide. However, the current push towards widespread integration of artificial intelligence (AI) models, including in E2EE systems, raises some serious security concerns.

This work performs a critical examination of the (in)compatibility of AI models and E2EE applications. We explore this on two fronts: (1) the integration of AI assistants within E2EE applications, and (2) the use of E2EE data for training AI models. 
We analyze the potential \emph{security implications} of each, and identify conflicts with the security guarantees of E2EE. Then, we analyze \emph{legal implications} of integrating AI models in E2EE applications, given how AI integration can undermine the confidentiality that E2EE promises. Finally, we offer a list of detailed \emph{recommendations} based on our technical and legal analyses, including: technical design choices that must be prioritized to uphold E2EE security; how service providers must accurately represent E2EE security; and best practices for the default behavior of AI features and for requesting user consent. We hope this paper catalyzes an informed conversation on the tensions that arise between the brisk deployment of AI and the security offered by E2EE, and guides the responsible development of new AI features.
\end{abstract}

\renewcommand{\baselinestretch}{0.1}\normalsize
\tableofcontents

\bigskip\noindent
{%
\textbf{Related Shorter Resources:} Blog-post summaries of this paper by some of the authors are available at the \href{https://nyudetail.substack.com/e2ee-ai-broken-promise}{NYU DeTaIL Lab Blog} \cite{nyu-detail-blog} and \href{https://www.techpolicy.press/can-bots-read-your-encrypted-messages-encryption-privacy-and-the-emerging-ai-dilemma}{Tech Policy Press} \cite{tpp-blog}. For examples of others' commentary on our work, see \href{https://blog.cryptographyengineering.com/2025/01/17/lets-talk-about-ai-and-end-to-end-encryption}{Matt Green's perspective} in his blog A Few Thoughts on Cryptographic Engineering \cite{matt-blog}, and \href{https://nyudatascience.medium.com/ai-assistants-in-encrypted-messaging-moving-too-fast-in-the-dark-8680ab195834}{Stephen Thomas's succinct coverage} on the NYU Center for Data Science blog \cite{nyu-cds-blog}.}
\renewcommand{\baselinestretch}{1}\normalsize

\section{Introduction}
\label{sec:intro}

Widespread adoption of end-to-end encryption (E2EE)\footnote{We use ``E2EE'' to mean either ``end-to-end encryption'' or ``end-to-end encrypted,'' depending on context.} has improved the confidentiality and integrity of data in various contexts, including messaging, video, and voice communication. Leading applications---such as iMessage~\cite{imessage}, WhatsApp~\cite{whatsapp}, and Signal~\cite{signal}---have integrated E2EE into their systems by default, bringing strong confidentiality and privacy to billions of users' communications. 
This trend appears poised to expand further, with an increasing number of services committing to the adoption of E2EE in order to ensure ``safer, more secure and private service'' to users~\cite{discord, messenger}. 

In an era where so many forms of communication have moved online, widespread E2EE has become an essential underpinning for the security of communications, providing critical protections for journalists, dissidents, and activists, as well as the everyday family and social communications of billions. 
As underscored by independent experts from the UN Office of the High Commissioner on Human Rights, strong encryption is essential to fundamental human liberties including free expression, the right to hold opinions, the right to privacy, and the right to assembly~\cite{ohchr}. Beyond individual privacy, the ubiquitous nature of E2EE offers \emph{systemic} protections for particular at-risk users, preventing their activities from appearing exceptional: widespread deployment of E2EE has made privacy the norm. This normalization of secure communications, in turn, makes it harder to distinguish whether individuals are engaged in frivolous or sensitive activities, and deters the targeting of individuals merely for using privacy technologies, resulting in a critical ``network effect''~\cite{WhittakerInterview} that creates the conditions necessary for journalists, human rights defenders, activists, and others to communicate securely with less fear of surveillance and repercussions---protections that a special-purpose platform just for top-secret communication would not be able to provide. These ``network effects'' currently protect privacy and freedom of expression for at-risk groups and everyday users around the world. 

The movement towards widespread E2EE communication stayed strong alongside the ``big data'' trends of recent decades: a recognition of the value of keeping private messaging data out of increasingly vast repositories of personal information, even as data was called the new oil by some~\cite{data-new-oil}. Privacy technologies \emph{gained} momentum as prominent examples of government and corporate intrusions on individual privacy came to light~\cite{looking-back-snowden,benner2016apple,aclu-e2ee}, and people became increasingly aware of their digital activity being tracked and profiled at large scale~\cite{pew2019privacy}. Those providing E2EE services were popularly seen as protectors of consumer privacy and freedom of speech \cite{nyt-tim-cook}. Both corporate and government surveillance were key issues, but corporations often stepped into the ``protector'' role and deployed E2EE technologies that would permit nobody other than the users participating in a communication to access private communication content: a decision that made business sense at the time~(e.g., \cite{benner2016apple}).

Following remarkable recent advances and an explosion of interest in large language models and generative artificial intelligence (AI) more broadly, however, we observe three trends that raise alarms for E2EE security. 

\emph{First, the way people interact with AI models is changing.} While they originally served as standalone applications, AI models are now increasingly incorporated into other everyday applications and throughout devices, including messaging applications, in the form of AI ``assistants.'' Interacting with these assistants is often baked into the user experience by default, made readily available as part of the application client (e.g., within a messaging app). Such integration creates new systemic data flows at scale between previously separate systems, and accordingly raises security and privacy considerations not limited to E2EE.

\emph{Second, high-quality training data is becoming scarce.} This has created a race between model developers, with tech companies under increasing pressure to tap any potential source of human-written content~\cite{villalobosposition}. That which is publicly accessible online has already been harnessed, so privately held data is naturally the next resort, and indeed, companies have been quietly changing their terms of service to enable training AI models on more and more of the user data they hold \cite{tan2024tos}---a practice about which the U.S. Federal Trade Commission has raised concerns~\cite{ftc2024tos}. At the time of writing, E2EE messaging data on major existing platforms remains off limits for AI training, but companies may be incentivized to reconsider these policies in the near future.
Apple and Samsung have very recently launched integrated AI features that can process E2EE content on their devices (see Section~\ref{sec:apple}), and other companies may be considering similar options. 

\emph{Third, recent trends in AI appear to have shifted business incentives.} Over the last decade, E2EE became a central part of the business model of online intermediaries. Business incentives to protect proprietary and user data were aligned with human rights protections. For example, in 2016, Apple and Facebook were lauded for their public resistance to government requests for access to E2EE data, including their refusal to comply with government requests to build systemic weaknesses into their encryption systems which would allow law enforcement access to messaging data~\cite{benner2016apple}.
However, in the last couple of years we have seen a shift towards prioritizing AI features as a component of modern applications, and a corresponding search for sources of data to power these, sometimes even at the cost of existing product features and profits~\cite{nyt-search-furor,forbes-customer-experience,arstechnica-profits}. At the same time, interestingly, research and reporting shows that consumers themselves can be much more skeptical of AI-powered products~\cite{kikuchi2023growing,cicek2024adverse}.

Against the backdrop of these three concerning trends, the motivating question of our paper is as follows:

\begin{center}
Is processing of E2EE content by integrated AI models compatible with end-to-end encryption? \\
\emph{(If so, to what extent and under what circumstances?)}
\end{center}

E2EE is a context in which confidentiality is the core guarantee above all others---confidentiality is the essence of the product. 
Though E2EE may be relatively unusual in this respect, our inquiry in this paper resonates with themes in the broader debate about security, privacy, and trust in AI. 

\paragraph{Summary of contributions}
We systematically analyze the above question from both a technological and a legal perspective, combining the expertise of an multidisciplinary team across cryptography/security, AI/LLMs, and technology law. 
Our contributions and recommendations illustrate that technological solutions and legal frameworks must work in tandem in order to achieve effective and enforceable safeguards. %

\bi{First}, we identify the key confidentiality and integrity properties provided by E2EE, for both \emph{individual} and \emph{systemic} privacy, which might be adversely impacted by feeding E2EE content into AI models (\S\ref{sec:defs}). This requires a novel conceptualization of key properties of E2EE, ranging from the strictly definitional, to realities of E2EE deployments and their divergences from the strict definitions, to essential properties that only hold when E2EE is used at scale, to auxiliary properties that are widely associated with E2EE. %

\bi{Second}, we examine a range of technical configurations for AI processing of E2EE data, taking into consideration the state of the art in cryptography and privacy technologies, as well as the latest technical developments in AI (\S\ref{sec:technical-implications}). We assess the capacity of each such technical configuration to uphold the guarantees of E2EE. We conclude that some configurations \emph{cannot} uphold these guarantees, while some others can.

\bi{Third}, we overview relevant areas of law, and provide an analysis of the circumstances under which E2EE service providers are likely to be able to offer AI features which use E2EE content, consistent with their legal obligations under current US and EU law (\S\S\ref{sec:background-legal}--\ref{sec:legal-implications}). We highlight areas of legal uncertainty, and discuss pending legal (or quasi-legal) processes.

\bi{Finally}, our analysis yields the four key recommendations below (\S\ref{sec:recommendations}; italic key terms defined in \S\ref{sec:defs:e2ee-terms}).
Our analysis indicates that there would be serious risks---security risks and legal risks---involved in diverging from these recommendations and thus undermining the established operation of the current E2EE ecosystem.

\begin{enumerate}
    \item \textbf{Training.} Using end-to-end encrypted content to train \emph{shared AI models} is not compatible with E2EE.

    \item \textbf{Processing.} Processing \emph{E2EE content} for AI features (such as inference or training) may be compatible with end-to-end encryption only if the following recommendations are upheld:
        \begin{enumerate}
            \item Prioritize \emph{endpoint-local} processing where possible.
            \item If processing E2EE content for \emph{non-endpoint-local} models, 
                \begin{enumerate}[label=(\roman*)]
                    \item No \emph{third party} can see or use any \emph{E2EE content}\footnote{Or any derivative of E2EE content (i.e., any non-trivial function of content). See Section~\ref{sec:defs:e2ee-terms} for full definitions.} without breaking encryption,\footnote{This may sound paradoxical; see Section~\ref{sec:technical-implications:private-inference} for detailed discussion of technical configurations that may achieve this condition.} and
                    \item A user's \emph{E2EE content} is exclusively used to fulfill that user's requests.
                \end{enumerate}
        \end{enumerate}

    \item \textbf{Disclosure.} Messaging providers should not make unqualified representations that they provide E2EE if the default for any conversation is that %
    \emph{E2EE content} is used (e.g., for AI inference or training) by any \emph{third party}.

    \item \textbf{Opt-in consent.} AI assistant features, if offered in E2EE systems, should generally be off by default and only activated via opt-in consent. Obtaining meaningful consent is complex, and requires careful consideration including but not limited to: scope and granularity of opt-in/out, ease and clarity of opt-in/out, group consent, and management of consent over time.

\end{enumerate}

\paragraph{Roadmap}
In Section~\ref{sec:background} (Background), we overview the properties of E2EE (\S\ref{sec:background:e2ee}), how AI assistants work (\S\ref{sec:background:ai}), how AI assistants can process E2EE data (\S\ref{sec:background:how_is_it_possible}), and some background on trusted hardware (\S\ref{sec:background:trusted_hardware}).
In Section~\ref{sec:apple}, we discuss three examples of recently deployed AI assistants that can process data from E2EE environments.
Section~\ref{sec:defs} provides definitions of key terms that we rely on throughout the paper (\S\ref{sec:defs:e2ee-terms}), including a taxonomy from ``strict E2EE'' to ``not E2EE'' (\S\ref{sec:buckets}). 

In Section~\ref{sec:technical-implications} (Technical Implications), we analyze and present a range of technical configurations in which AI assistants might process E2EE data, and associated security/confidentiality implications. In Sections~\ref{sec:background-consent}--\ref{sec:sociotechnical-implications} (Legal Implications), we overview different areas of US and EU law and how they may constrain the use of E2EE data as input for AI models. 
In Section~\ref{sec:recommendations}, we offer our detailed recommendations on how and when messaging data can be fed into AI models consistent with E2EE, taking into careful consideration both technical possiblities and limitations, and existing legal frameworks.
Section~\ref{sec:discussion} provides discussion
and Section~\ref{sec:conclusion} concludes
(and includes a detailed note on author contributions).

\section{Background I: Encryption, AI, and Trusted Hardware}
\label{sec:background}

This section covers technical background: end-to-end encryption (Section~\ref{sec:background:e2ee}), AI assistants (Section~\ref{sec:background:ai}), how it is possible for AI assistants to process E2EE content (Section~\ref{sec:background:how_is_it_possible}), and trusted hardware (Section~\ref{sec:background:trusted_hardware}).
Section~\ref{sec:apple} gives a brief overview of three deployed systems in which AI assistants can process E2EE data: Apple Intelligence, Samsung's new AI-integrated devices, and the integration of MetaAI with WhatsApp.

For readers knowledgeable about E2EE, TEEs, and AI, we suggest skipping to Sections~\ref{sec:background:trusted_hardware:tees_v_e2ee} and \ref{sec:apple}.

\subsection{Background on End-to-End Encryption (E2EE)}
\label{sec:background:e2ee}
Online messaging systems that allow communication between users (such as iMessage, WhatsApp, and Signal) are generally intermediaries to every communication. Although the user experience may make it seem like messages go directly from a sender $S$ to a recipient $R$, in reality every message goes from $S$ to the platform $P$ and then from $P$ to $R$.
This means that, absent any additional guarantees, the platform has access to all communications of users on the platform.

End-to-end encryption (E2EE) is a standard of security for communication in which only the sender and an intended recipient can read communications between them.
In particular, even the platform $P$ cannot read these communications despite still acting as the intermediary for communication: that is, communications still go from $S$ to $P$ then $P$ to $R$.

The key tool used in E2EE is \emph{encryption}.
Encryption is a method for a sender $S$ to ``scramble'' their communications in a manner which is easy for their intended recipient $R$ (and $S$) to unscramble, but practically impossible for anyone else.
Using encryption, $S$ will scramble (or \emph{encrypt}) each message for $R$ before sending it to platform $P$, which only handles messages in scrambled form (and does not have the ability to unscramble messages).
When $R$ receives a scrambled message from $S$ via $P$, they are able to unscramble the message (or \emph{decrypt} it) and read the original message that $S$ wrote.

The additional security that E2EE provides is an important protection for the privacy of the communicating users, as well as a key protection against misuse of the message contents by or at the platform $P$: e.g., the messaging data is safely inaccessible if $P$ suffers a data breach, if $P$ is asked to share data with other organizations (e.g., business partners or the government), or if an insider at $P$ wants to read the data.

We conclude this section with some terminology and follow it with a slightly more detailed and technically accurate explanation of how E2EE works.

\paragraph{Plaintext and Ciphertext} Message content that is unencrypted is described as \emph{plaintext}.
Normal, readable messages that you type into iMessage, Signal, or WhatsApp are plaintexts.
Despite the name, plaintexts can include any type of content: audio files, images, spreadsheets, and any other digital information.
Encryption provide a technique to convert (or scramble) plaintexts into \emph{ciphertexts}, which are unreadable except to the designated recipient (and the sender who wrote it).\footnote{The terms \emph{plaintext} and \emph{ciphertext} can each be used either as a noun or as an adjective.}

\paragraph{Secret Key}
Each user in an E2EE system has a \emph{secret key}, a secret piece of information that should be known only to them (much like a password).
A user's secret key is necessary to decrypt messages intended for them.
Secret keys are created and stored on users' devices: platforms (and other users) do not have access to them.

\paragraph{Encryption and Decryption}
The process of converting a plaintext into a ciphertext is called \emph{encryption}, and the process of converting a ciphertext back into a plaintext is called \emph{decryption}.
Only the intended recipient $R$ should be able to perform decryption to recover plaintexts from ciphertexts; accordingly, decryption requires the use of some secret information which only $R$ knows, namely, $R$'s \emph{secret key}.

\paragraph{Putting it together}
Equipped with more formal vocabulary, we restate the basic mechanism of E2EE.
In E2EE messaging, each sending user $S$'s device encrypts each plaintext message, then sends the resulting ciphertext to the platform $P$.
The platform $P$ passes the ciphertext on to the recipient user $R$'s device, which uses its secret key to decrypt the ciphertext and get back the original plaintext message that $S$ wrote.

\begin{figure}
    \centering
    \includegraphics[width=0.8\linewidth]{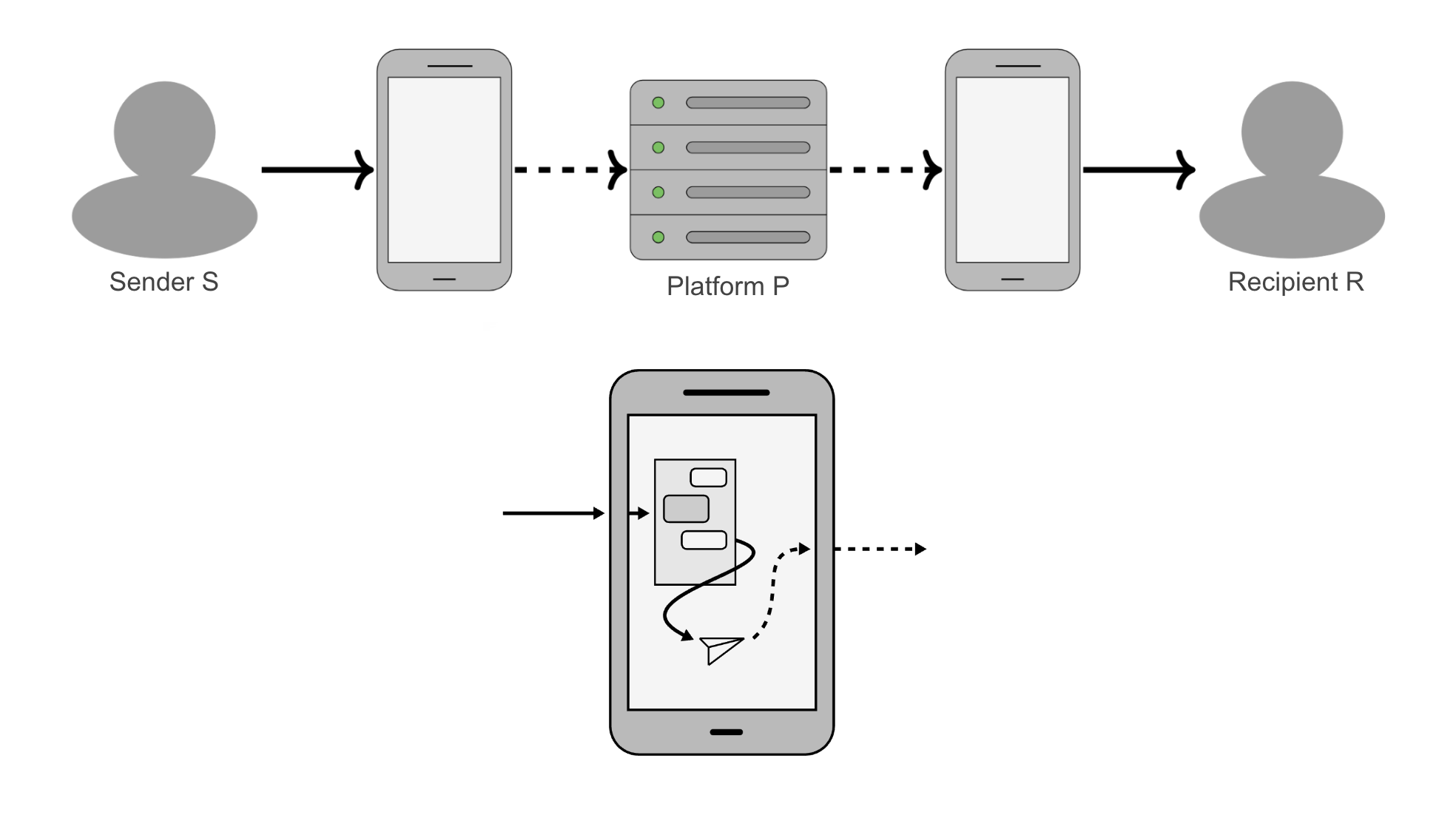}
    \caption{A sender $S$ and receiver $R$ communicate using an end-to-end encrypted application hosted by a company (middle). Solid lines represent plaintexts, and dashed lines represent ciphertexts. $S$ and $R$ can read their messages on their devices; however, while a message is ``in transit'' between their devices, it is encrypted so that it is not readable to the intermediary platform $P$ handling it on its servers (or indeed to anyone but the intended recipient, whether a network eavesdropper, an employee at $P$, a hacker who compromises $P$, or a wrong recipient to whom the message was delivered by mistake). \betty{not clear that the below is a more detailed depiction of the two devices above, maybe add some zoom out lines, or at least has to be explained in the caption} } 
    \label{fig:e2ee_sketch}
\end{figure}

\subsubsection{The E2EE Confidentiality Guarantee}\label{sec:e2ee-confidentiality}%
\label{sec:background:e2ee:e2ee-confidentiality}

The core privacy guarantee of an E2EE system is called \emph{confidentiality}.\footnote{We will hereafter use the term ``E2EE confidentiality'' in this paper to stress when we are referring to this particular technical property, and not the common use of the word.}
Informally, E2EE confidentiality states that plaintext messages handled by an E2EE system can only be seen by the sender and intended recipients: no one else, including law enforcement or the service provider, can learn anything about the plaintext messages~\cite{e2ee-definition-rfc}. Importantly, this guarantee is stronger than saying that nobody else can \emph{learn the plaintext messages}: E2EE confidentiality requires that nothing can be learned about the content of an E2EE message, including the message itself, but also any derivative of the message content.\footnote{The formal mathematical definition of security of encryption states that given the message length in bits, no computational adversary can guess any function of the message content with success non-negligibly better than random guessing. \label{fn:formal-e2ee-security}}
Thus, E2EE confidentiality means that anyone besides the sender and intended recipient must not be able to learn, for example, what letter of the alphabet a message begins with, or whether the message contains a given word.
See~\cite{wired_e2ee_def, hale2022end, e2ee-definition-rfc, scheffler2023sokcontentmoderationendtoend} for more detailed definitions of end-to-end encryption.

E2EE confidentiality (or E2EE security) refers to a specific technical definition of confidentiality for messages, based on encryption. It is a term of art with a precise meaning, and does not ensure perfect ``confidentiality,'' ``privacy,'' or ``security'' for every colloquial meaning of the terms. In general use, these terms are very broad and mean different things in different contexts.\footnote{For example, E2EE technologies do not prevent your friend from choosing to disclose the sensitive information you shared with them to others, even though this might seem to ``violate confidentiality'' or ``violate your privacy'' in a broad sense. Indeed, no technology can prevent this.\label{fn:friend-secret} See also Section~\ref{sec:background-legal} for more discussion.}

\subsubsection{Auxiliary Security Properties of E2EE Systems}
\label{sec:background:e2ee:auxiliary-e2ee-properties}
Besides the core guarantees mentioned so far in this section, there are additional (cryptographic) security properties that E2EE applications often provide in practice.
While distinct from basic E2EE confidentiality, these properties have become very commonly associated with, or expected of, modern E2EE applications. Notable examples include:
\begin{newitemize}
    \item \emph{Integrity}~\cite{unger2015sok}: Plaintext messages cannot be modified during transit in undetectable ways.
    \item \emph{Forward secrecy}~\cite{green2015forward}: An adversary who has compromised the communication channel (e.g., by stealing the victims' secret session key) cannot recover messages that were transmitted before the session was compromised.
    \item \emph{Post-compromise security}~\cite{cohn2016post}: An adversary that has compromised the communicate channel cannot recover messages that are transmitted after the protocol ``heals'' itself (e.g., after session keys are rotated).
    \item \emph{Deniability}~\cite{canetti1997deniable}:  No party can prove that a message was authored by any particular user of the system, even if the prover was one of the intended recipients of the message.
    \item \emph{Unlinkability}~\cite{unger2015sok}: Knowledge of the fact that a user authored a particular message reveals no information about whether they authored other messages.
\end{newitemize}

State-of-the-art E2EE messaging protocols, such as the Signal protocol~\cite{signal}, meet these (and many additional) security properties.

\subsubsection{Systemic Properties of E2EE Systems}
\label{sec:background:e2ee:systemic-e2ee-properties}
Even if an E2EE application provides strong technical privacy guarantees (including and beyond E2EE confidentiality),
the strength of the privacy guarantee it offers will also depend on its usage in practice.
Messaging applications that are widely used offer additional protection because their use is \textit{unremarkable}, and hence, the use of the application in itself does not draw attention.
In economic terms, this is a \emph{network effect} where the (privacy-related) utility of the application for any given user depends upon other users.

For example, suppose that a particular E2EE messaging app $A$ is known to have a userbase which disproportionately represents a particular political community.
Even if $A$ provides perfect E2EE confidentiality, knowing that a particular person uses $A$ can divulge their political views, if only by association.
This type of example motivates the intuition that to realize the full privacy potential that E2EE can provide, the application should be commonplace and its use should be unremarkable in the relevant context.
This ``unremarkableness'' is not a technical property: it is instead an emergent social and contextual property of the system's use as a whole considered over the entire userbase (relevant to a context of use\footnote{E.g, widespread use in country $X$ may not be helpful for a user in country $Y$.}).
Existing security and privacy literature has discussed similar network effects in privacy technologies not limited to encrypted messaging, and noted their importance to engineering effective privacy technologies (e.g., Tor)~\cite{DBLP:conf/weis/DingledineM06}.
This key privacy property of E2EE systems, which cannot be achieved by technical means alone, \emph{is} attained by popular E2EE applications such as iMessage and WhatsApp today---namely, the network effect gained from wide usage such that usage in itself is unremarkable.

In summary, there is a key privacy property of E2EE systems which cannot be achieved by technical means alone, but \emph{is} attained by popular E2EE applications such as iMessage and WhatsApp today---namely, the network effect gained from wide usage, so that the use of the system in itself is unremarkable.

\subsubsection{Necessary Conditions for E2EE Confidentiality}\label{subsubsec:e2ee-assumptions}

End-to-end encryption provides its promised confidentiality guarantee only under certain conditions: namely, when (A) the underlying encryption method is secure and (B) that encryption method is securely implemented in the application being used.

\paragraph{Cryptographic assumptions}
Every practical encryption scheme relies upon a \emph{hardness assumption}: an assumption that a particular mathematical problem is difficult for computers to solve.
The hardness of the mathematical problems underlying today's widely used encryption schemes has held strong in the face of heavy scrutiny over decades by mathematicians, computer scientists, and others, leading to a strong community belief that it is acceptable to use cryptography that relies on these assumptions in security-critical contexts.
When new encryption schemes that rely on new, less-tested assumptions are proposed, they are not considered suitable for deployment until they have been vetted in a multi-year process of scrutiny by mathematicians and computer scientists worldwide.

\paragraph{Practical assumptions}
Even if the math underlying an encryption scheme is perfect in theory, the process of deploying the encryption scheme in software or hardware introduces avenues for security failures, and the use of the encryption scheme by fallible humans in practice introduces further potential for security failures.
In order for E2EE confidentiality to hold: (1) the code implementing the E2EE application must be correct, (2) the hardware device that the application runs on must be secure, and (3) the users using the application must use it correctly and take adequate security precautions.\footnote{In fact, these are essential considerations for the proper functioning of any software product, not limited to encryption.}
Undermining any one of these three could undermine E2EE confidentiality: for example, (1) the software fails to perform the encryption correctly; (2) the hardware device is infected with malware; or (3) a user fails to keep their secret key secret.
Errors can be introduced in myriad ways: through simple mischance, human mistake, or intentional sabotage.

Typically, in current practice, a messaging application is implemented by an intermediary platform that provides the associated messaging service.
Thus, the party with the most control over and visibility into the implementation process is the platform.
Users therefore rely on platforms for (1), i.e., that the code implements E2EE correctly.
Users rely on device manufacturers (and their suppliers) for (2), i.e., that the device running the app is secure. 
Finally, the users themselves determine (3), i.e., their usage of the application; that said, users often lack the expertise to confidently take adequate security precautions.

\subsubsection{E2EE Messaging Features and Usability}
\label{sec:background:e2ee:usability}

Many academic definitions of E2EE messaging consider a simple model where the only activity that occurs is literal messaging---sending messages between senders and receivers and within group chats.
However, modern messaging apps are more feature-rich, and the ways in which such additional features process message contents---sometimes in tension with E2EE confidentiality---has long been a subject of debate.
Some common features are, strictly speaking, in violation of E2EE confidentiality: for example, \textit{link previewing}, where the app displays a ``preview'' of a link contained in a message, technically violates E2EE confidentiality because a third party processes the link in order to fetch the preview.
Furthermore, some common features of E2EE applications \emph{enhance} confidentiality despite not being part of the basic definition of E2EE: for example, \textit{user blocking} helps protect users from harassment and spam, but has no direct relationship to E2EE confidentiality (or any of the other cryptographic security properties mentioned in Section~\ref{sec:background:e2ee:auxiliary-e2ee-properties}).\footnote{Other common features with potential consequences for confidentiality and E2EE confidentiality under some implementations include: importing contacts, autocorrect, and accessibility features such as text-to-voice.}

In short, any comprehensive treatment of E2EE messaging applications must consider the interaction of commonplace messaging features with the system.
As discussed in Section~\ref{sec:background:e2ee:systemic-e2ee-properties}, an E2EE application requires a sufficiently large and diverse user base to promise privacy in practice, and it is therefore pragmatically important for E2EE messaging apps to incorporate common usability features of messaging apps if doing so could make or break users' willingness to choose the app.
Of course, this is not an argument for a kitchen-sink approach to messaging app development: some features may be too incompatible with E2EE, or can at least be made more palatable by choosing certain implementations over others.
For example, Signal has link previewing, but also offers users the ability to disable link previewing in the settings.

This complicated relationship between messaging features, usability, and privacy in E2EE apps has led to practical controversy as well as academic debate.
For example, Apple offers a feature where users can store encrypted backups of their iPhone data on Apple's cloud storage---a standard convenience.
In February 2025, reports emerged that the UK government has ordered Apple to grant it backdoor access to these encrypted backups~\cite{apple-uk-backups}.
In this case, Apple providing (encrypted) backups---providing utility to the end user---also created a new resource for third parties (here, law enforcement) to target to gain access to end-to-end encrypted user communications, potentially without users' knowledge and without breaking encryption.

While it is out of our scope to fully characterize the wide spectrum of messaging app features and their privacy implications, it is useful to view AI assistants in the context of a broader constellation of other privacy-impacting messaging app features which have been met with controversy and debate.

\subsection{Background on AI Assistants}
\label{sec:background:ai}

AI assistants are programs designed to interpret everyday language and perform computational tasks. Today's AI assistants, such as ChatGPT, Gemini, and Llama, are able to handle a wide range of tasks, including text analysis, content creation, code generation, language translation, and more. At the core, these technologies are statistical models---programs trained on data to identify patterns. Specifically, AI assistants are typically built using large language models (LLMs), which are trained on vast datasets of text to generate outputs based on predictions of the next word. Due to the size of their training datasets and sophisticated architectures, LLMs can handle complex, previously unseen user inputs (\emph{queries}) and provide contextually relevant responses.

\subsubsection{Training and Inference}

AI assistants undergo two key phases: \emph{training}, when the model is built to identify patterns, relationships, and rules; and \emph{inference}, when real-time responses are generated based on user inputs. Training itself often involves multiple stages, typically referred to as \emph{pretraining} and \emph{fine-tuning}. In pretraining, the model takes in large-scale general datasets to learn structures and patterns in the data. Fine-tuning then refines this pretrained model by training it on additional, more specialized datasets, improving its accuracy and relevance in particular contexts. We describe these processes in further detail below.

\paragraph{Training} Models have configurable \emph{parameters} which are numerical values that determine how outputs are generated. These parameters are used in calculations to predict outputs when the model is queried with an input. During training, the model processes large datasets---such as books, websites, code repositories, and conversational datasets in the case of LLMs---which iteratively adjusts (or ``tune'') the parameters, capturing more and more refined patterns in the data with repeated passes. The finalized set of parameters constitutes the trained model, which can then be \emph{prompted} with inputs (e.g., questions) to generate outputs (e.g., answers). Models at the scale of commercial AI assistants contain billions or even trillions of parameters, trained over the course of months on high-performance computers.

\paragraph{Inference} Once trained, the model can be used for inference, where it outputs real-time responses to user inputs. When a query is inputted to a model, it is processed with the model parameters to calculate an output. While inference is less resource-intensive than training, it still requires significant computational power and is typically handled by centralized servers.

\begin{figure}[ht!]
    \centering
    \includegraphics[width=0.55\linewidth]{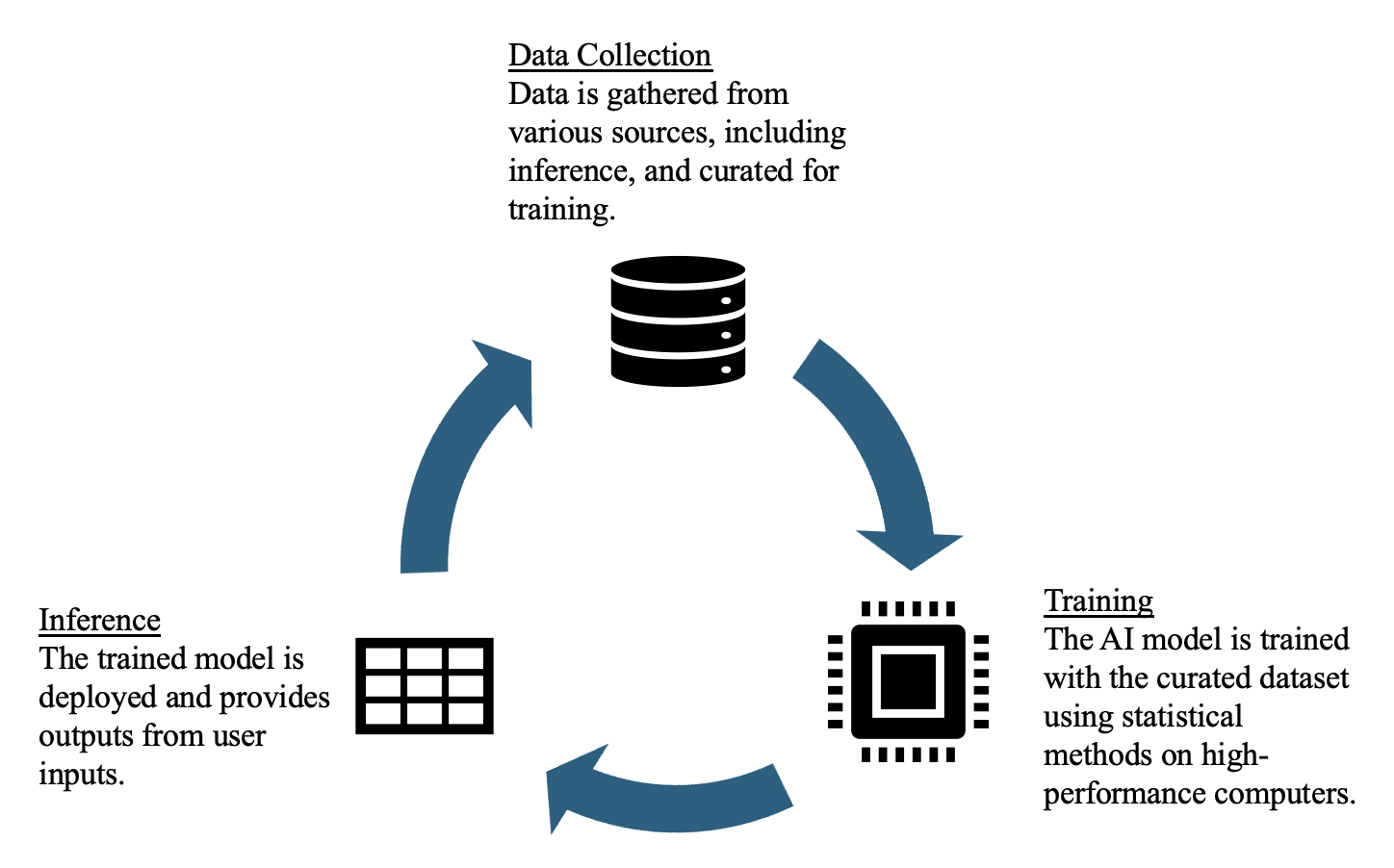}
    \caption{Inference and training of AI assistants happen continuously as a feedback loop. As users continue to query the AI assistant, this generates data that is stored and can be used to further train and improve the model.}
    \label{fig:ai-training-loop}
\end{figure}

Deployed models undergo continuous improvement through updates that often involve additional training on more diverse or recent datasets. Some approaches for this include reinforcement learning with human feedback, where collected user feedback is used to improve response quality \cite{ouyang2022training}; patches to address issues such as specific problematic outputs, and model refinement with new techniques and architectures. As such, inference and training often overlap in a \emph{feedback loop}, where data collected during inference is fed back into the training pipeline to fine-tune the model further. This process is shown in Figure \ref{fig:ai-training-loop}.

\subsubsection{Data Collection}
Once the AI assistant is deployed, various types of data from the user interactions with the assistant may be collected. This can include \emph{chat logs}, which record user inputs (text, voice, or image data) and the system's responses. These historical records, even if kept by the platform, may or may not be visible to users.\footnote{Search/query history can be surprisingly revealing when aggregated over time, involving family concerns, health issues, religious beliefs, and more. Research has shown that users can be reluctant to disclose all their search history even for better personalized search results \cite{PanjwaniSSJ13}.} Additionally, \emph{metadata}, such as timestamps, device type, language, and geographic region, as well as user preferences and settings, may also be collected and stored by the platform. This data serves multiple purposes, which can be broadly categorized as follows.
\begin{enumerate}
    \item Ease of use: The data is used to enable certain features, such as maintaining conversation histories or continuity within a session/across sessions, for an individual user.
    \item Personalization: The data is used to identify usage patterns, such as query frequency and commonly accessed features, to tailor the system's functionality for an individual user. 
    \item Analytics: The data and usage patterns is aggregated across users to steer further improvements and developments of the model. 
    \item Model refinement: The data is used as training data to fine-tune the model and continuously improve its performance. 
\end{enumerate}

Collecting user data is commonly enabled by default with collected user data transmitted and stored on centralized cloud servers. In some cases, some or all user data may remain local on the user's device, eliminating external transmission.

\subsubsection{Architecture of AI Assistants in Messaging Platforms}
Next, we overview the architecture of AI assistants integrated in messaging platforms. The architecture is how the application is structured technically, which has implications on how and where data is processed.  First, it is important to distinguish between the messaging application, the AI assistant, and the model. The messaging application is the application installed on the user's device which facilitates communication between users and provides the interface for interacting with the AI assistant, e.g., WhatsApp, Messenger, and Signal. The AI assistant is the system that performs tasks such as answering queries, generating responses, or providing recommendations, e.g., Meta AI, which is built into WhatsApp and Messenger. At its core is the model---in this case, an LLM---the underlying computational engine that processes inputs and generates outputs based on patterns learned during training.

When the messaging application gets an input from the user, this input is sent to the model in the form of a \emph{request}. Once the request is processed and an output has been generated by the model, the output is sent back to the platform as a \emph{response}. As such, while the messaging application itself will be installed on the user's device, the model need not also be directly on the device; network connections can facilitate this communication between the application and model.

\begin{figure}[h]
    \centering
    \includegraphics[width=0.9\linewidth]{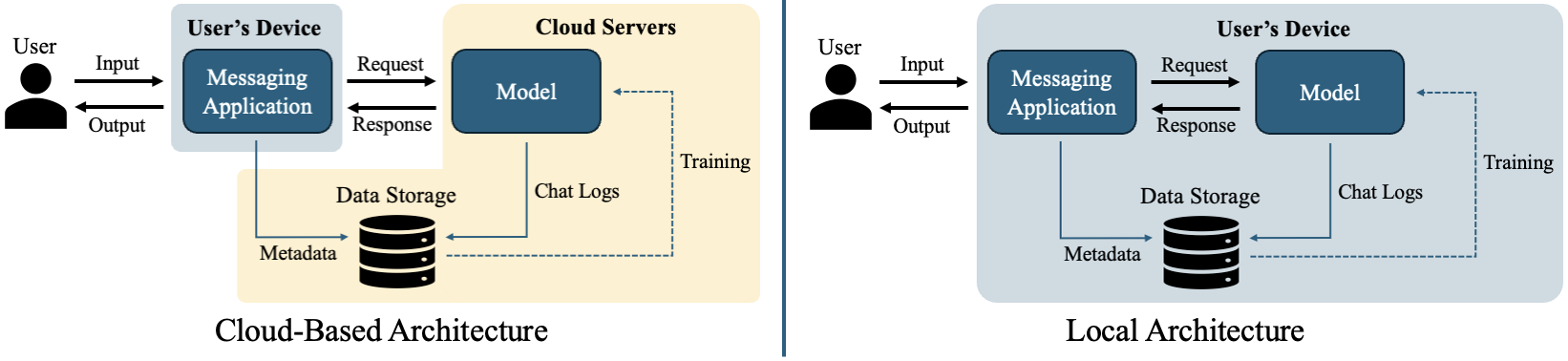} 
    \caption{Generating an output from a user's input, differentiating between what is on-device and on cloud servers between a cloud-based and local architecture. 
    }
    \label{fig:background-archs}
\end{figure}

We differentiate between two types of architectures: \emph{cloud-based} and \emph{local}, illustrated in Figure \ref{fig:background-archs}.

\paragraph{Cloud-Based} Cloud servers are designed to handle large computation and storage demands. Given the demands of an LLM---recall that there may be trillions of parameters to be processed simultaneously, requiring very high levels of computation power as well as memory to store the parameters and data---typically, models are hosted on cloud servers. Cloud servers can carry thousands of high-performance computers (Graphics Processing Units or Tensor Processing Units), exceeding the capabilities of a typical laptop by many magnitudes.

\paragraph{Local} In limited scenarios, computations can occur \emph{locally}---that is, directly on the user's device using lightweight versions of the model or specific components. Local processing ensures that data does not leave the device, but is only possible with smaller, less powerful models due to reduced processing power and storage capacity of local hardware.

\subsubsection{Standalone vs. Integrated AI Features in E2EE Applications}
\label{sssec:standalone-vs-integrated}

AI assistants are commonly available in major E2EE applications, as a standalone tool \emph{separate} from user conversations (and any other E2EE content). For example, an E2EE messaging application might also include a search bar where users can type in queries to search the web and/or send to AI models. 

In general, we do not consider such standalone tools to be \emph{integrated} with E2EE applications in the way we are concerned about in this paper. Standalone AI features separated from E2EE content generally do not raise the security concerns that we argue \emph{do} arise with integrated AI features, unless providers create the impression that these AI features are themselves E2EE. We consider standalone features out of our scope. (Section~\ref{subsec:ai_integration} provides further discussion with practical examples based on Meta AI features in WhatsApp.)

Our focus is, rather, on integrated AI models and their capacity to process E2EE content.
Integrating AI assistants as a feature within E2EE conversations would often require the AI assistant to process E2EE content: for example, to allow users to invoke the AI assistant within an E2EE conversation or to provide contextually relevant support, responses, or summaries within E2EE conversations.

\subsection{How Is It Technically Possible for E2EE Data to Be Fed Into AI Assistants?}
\label{sec:background:how_is_it_possible}

The definition of E2EE may make it sound like it should be impossible to feed E2EE content into a (cloud-based) AI assistant. %
How can an AI model have access to plaintext data, when E2EE guarantees that only the sender and recipient can access E2EE content in plaintext? 

Certainly, the cloud-based service provider has access to the encrypted message---the ciphertext---and could input that ciphertext to an AI assistant. In this case, however, the AI assistant would not be able to read any meaningful information from the ciphertext, and thus would not be able to produce a useful answer, either.\footnote{This is mathematically guaranteed by the formal definition of secure encryption.}

The current discourse around using E2EE data in AI assistants, therefore, is focused primarily on the idea of feeding readable \emph{plaintext} versions of E2EE messaging data into AI assistants. Since only the sender and recipient devices in an E2EE system can read the plaintext messages, the sender or recipient devices would be the only ones who could provide the plaintext data to the AI assistants. This possibility appears to be primary focus of the ongoing discourse: that users' devices would be set up to send plaintext copies of E2EE content to cloud-based AI assistants for processing.

Nothing about E2EE technically prevents this from happening---indeed, no technical measure would both be able to allow senders and recipients to read their own messages, and prevent them from showing them to anyone else. This is a bit like no matter how secretively you encode and deliver your hand-written letter personally to your friend, 
nothing you can do can stop your friend from showing the decoded letter to other people once they have it. (This analogy falls short after a point. In the letter-writing example, your friend would likely have to take an affirmative, intentional action to show the letter to others; but in the encrypted messaging case, users' devices and applications can be set up by service providers to \emph{automatically} send plaintext copies of messages to third parties without any action being taken by users.)

\bigskip

\bigskip

\paragraph{But then, what is the point of end-to-end encryption?}
We have just explained that E2EE is a technology that allows transmission of messages in encrypted form so that nobody but the sender and intended recipient can read them, not even intermediary platforms. We have also explained that E2EE does not stop the sender and recipient devices from sending plaintext copies of the messages to anybody --- and that devices can do this in a way that is very hard to detect for average users. 

Then, \textit{is there any point to using E2EE?} After all, our devices could be routinely sending plaintext copies of all our messages \emph{to the messaging platforms themselves}, and E2EE would not stop this --- yet such a practice would undermine precisely the benefit that E2EE purports to provide. %

E2EE \emph{does} still have a point, for a number of reasons. First, it provides important protections when service providers \emph{choose} not to be able to read your messages, and implement E2EE in order to achieve this goal. Such a choice might be motivated by various reasons: it might provide a competitive edge, provide protection in the event of data breaches, or reduce compliance costs for companies. Once such a choice is made, the platform moreover has a reputational interest in making sure E2EE is achieving the stated goal that the platform not be able to read messages. Second, E2EE provides stronger protections in the case of popular messaging platforms that are widely used and therefore widely scrutinized by technical experts, including security and privacy experts. While subversions like sending plaintext copies of messages to the platform would be hard to detect for an average user, they would be much more easily detected under such scrutiny by experts --- and from the experts' scrutiny, average users benefit too. Third, E2EE provides even stronger protections in the case of popular messaging platforms that choose to openly disclose the details of how their system and security features work, for similar reasons. Balsa et al.~\cite{BalsaNP22} provide more detailed discussion of these and related issues around platforms implementing E2EE and other cryptography.

\subsection{Background on Trusted Execution Environments (TEEs)}
\label{sec:background:trusted_hardware}
\emph{Trusted hardware} refers to physical devices that are designed to ensure the confidentiality and integrity of computations, even if the owner of the hardware may be compromised. Trusted hardware is designed to create a secure area of computation, in which no information about the computation is released outside of the trusted hardware during execution of code, and that code will run exactly as specified. There are many types of technologies that fall under the broad umbrella of trusted hardware---such as hardware security modules (HSMs), trusted platform modules (TPMs), and trusted execution environments (TEEs)---which offer different capabilities. Of these technologies, TEEs are the most relevant for the implementation of AI functionalities, as they are designed to run  arbitrary programs; other technologies currently have more limited, bespoke use cases, such as key generation and storage. We thus focus on TEEs from here on.

We can conceptualize a TEE as a physical ``lockbox'', which is designed to execute arbitrary code such that no outside entities (including, for example, the operating system, other device processes, and third parties) can observe nor manipulate the state of the code as it is running. During initial setup of the TEE, a computer program---call it $\Pi$---is loaded inside the TEE. The TEE then has a single input/output interface, through it which receives inputs $x$ from outside the TEE, computes the program $\Pi$ on the inputs, and outputs the result $\Pi(x)$ from the TEE. During execution of $\Pi$, the TEE maintains some internal ephemeral state in memory, which stores temporary values during computation. TEEs are also often equipped with persistent storage, on which they can store data longer-term. See~\cite{pass2017formal} for a formal treatment of TEE abstractions.

\paragraph{TEE security goals} TEEs are intended to provide two core security guarantees. The first is \emph{confidentiality}, which means that the internal state of the TEE should not be visible to any parties during execution of $\Pi$; only $x$ and $\Pi(x)$ are revealed to third parties. Similarly, the persistent storage of the TEE should always be kept secret. The second goal of TEEs is \emph{integrity}, which means that $\Pi$ should be the exact program run by the TEE, which cannot be undetectably tampered with nor replaced by another program. Besides these two core guarantees, some TEE implementations have a third security goal, \emph{remote attestation}, which means that the TEE should be able provide convincing evidence that a given program $\Pi$ is the code that is loaded inside the TEE. Critically, these guarantees should hold even in the presence of a compromised TEE owner (e.g., a cloud provider): using a number of hardware-level protections, TEEs are designed to be tamper-resistant (for example, TEE chips often have physical seals that are designed to detect attempts to open them, upon which the TEE's storage is immediately erased). %

The goal of TEE design is to enable outsourcing of computations over user data to potentially untrusted (but more computationally powerful) environments, by loading the program inside a server-side TEE, and processing user data inside of it. As described so far, the inputs to and outputs from the TEE can be seen by the owner of the TEE. However, inputs and outputs can be hidden from the TEE owner with an additional step: encrypt inputs and outputs so that they can only be decrypted using secret key that is stored inside the TEE. 
Then, computations inside the TEE can still happen in plaintext, within the ``shield'' of the hardware protections provided by the TEE.

\paragraph{TEE performance}
Modern TEEs (such as Intel SGX~\cite{costan2016intel} and AMD SEV~\cite{kaplan2016amd}) are capable of supporting complex computations with practical overheads. For example, TEE-based security has been deployed in Signal's contact discovery mechanism~\cite{signalcontactdiscovery}, and even entire blockchains ~\cite{oasis}. However, efficiency is not the only engineering consideration for TEEs: their design of strict  isolation imposes important limitations on the programs that TEEs can run, since they cannot have non-trivial interfaces to the outside world. So, for example, TEEs generally cannot make Internet requests, and cannot rely on external sources of state.

\subsubsection{Necessary conditions for TEE security}\label{sec:tee-assumptions}
The security guarantees of TEEs naturally rely on the correct implementation of the TEE, and that the hardware has not been tampered with at any point in the TEE's life cycle (e.g., tamper-proof seals have not been removed, backdoors have not been inserted, etc). This requires trust in various actors across the hardware supply chain, from manufacturing, to assembly, to deployment. Further, users typically do not
handle TEEs directly; rather, they are embedded in devices
or handled server-side. As such, users rely on the application or device provider that a TEE is in fact being used for any given application.. While some TEEs have additional mechanisms to attest to their usage for certain computations (e.g., via their attestation service), this generally needs to leverage some trusted authority, such as the hardware provider.

Even if the hardware is correctly implemented according to specification, the security of TEEs additionally relies on the specification itself being free of vulnerabilities. Unfortunately, there is a history of practical attacks on various TEE implementations~\cite{van2024sok, fei2021security}. These attacks range across a variety of threat models---including adversaries that have physical access to the TEE, can launch co-located TEEs with arbitrary programs, control the system's boot process, etc.---and violate the confidentiality and integrity of the TEE, i.e., recover sensitive runtime state or tamper with the data in the TEE. Some of the attacks have been mitigated since their discovery; others have not, and cannot be fully mitigated with known techniques~\cite{van2024sok}. Notable examples of attack vectors include physical and micro-architectural side-channels, which exploit shared caches~\cite{brasser2017software}, branch predictors~\cite{lee2017inferring}, page faults~\cite{xu2015controlled}, and transient execution~\cite{van2020lvi}.

\subsubsection{TEEs vs E2EE}
\label{sec:background:trusted_hardware:tees_v_e2ee}
While TEEs and encryption have similar security goals, there are some critical differences between the two technologies, which we review in this section. We first explain why they are not interchangeable security solutions, and then why TEEs introduce additional security concerns not present in the encryption context.

\paragraph{TEEs and encryption provide different security guarantees} E2EE security and TEE security each rely on a number of conditions holding true, as summarized in Section~\ref{sec:tee-assumptions} just above (for TEEs) and in Section~\ref{subsubsec:e2ee-assumptions} (for E2EE). The practical guarantees provided by each technology depend on both the precise security goals they are designed to achieve, and the conditions (or the \emph{threat model}) under which they are designed to achieve them. While TEEs and E2EE have overlapping security goals, the conditions necessary for E2EE security and for TEE security are \emph{orthogonal}---qualitatively different and incomparable. 

The security of E2EE and TEEs rely on two broad assumption types: (1) specification correctness, and (2) implementation correctness. The former states that the design of the technology/protocol is secure, while the latter states that a particular instance of the technology/protocol is correctly implemented according to the specification.
TEEs and E2EE have important differences for both assumption types. Assumption type (1) generally hinges on a computational hardness assumption for E2EE; and on a robust hardware design for TEEs. On the other hand, assumption type (2) hinges on the security of software and hardware implementations for E2EE and TEEs, respectively. These entail a completely different set of considerations between E2EE and TEEs, since the software and hardware supply chains involve different entities, vendors, and processes.

Since the security of TEEs and E2EE depend on orthogonal considerations, their resulting security guarantees are \emph{incomparable} in a technical sense. That is, the precise type of security you get, and the conditions under which you get it, are different and not directly comparable.
TEEs rely upon a distinct set of conditions for their security from E2EE, and therefore cannot be substituted into an E2EE protocol without acknowledging that the nature of the security offered has changed, away from E2EE security.
Thus, again, while the security goals are similar, they are not interchangeable technologies.

\paragraph{TEEs raise certain security concerns that encryption does not} 
The difference in the type of security offered by TEEs and by encryption mean that depending on the context, one may be more suitable for use than the other. That said, when considering whether to use TEEs or encryption for a given application that might admit either, the former raise certain kinds of security concerns not present for the latter. Widely used encryption schemes today have largely withstood decades of scrutiny without critical confidentiality breaks; indeed, new encryption schemes that have not yet been subject to such scrutiny are widely considered unsuitable for use in security applications. Our understanding and scrutiny of TEE security is comparatively young, and the field has seen and continues to see a stream of discoveries of security flaws with impacts on deployed systems
(e.g. \cite{van2024sok, fei2021security}); at the same time, TEEs enable secure computation in promising new applications, which would not be practical based on cryptography alone.

\subsection{Deployments in Practice: Apple, Samsung, and Meta AI}
\label{sec:apple}

In this section, we overview three deployed examples of AI assistants that may process E2EE data: Apple Intelligence, Samsung's Galaxy AI, and Meta AI in WhatsApp. We apply our recommendations to the case of Apple Intelligence in Section~\ref{sec:recommendations:apple-intelligence}, as we found that the documentation currently available on Apple Intelligence supports a more detailed analysis than for the other two products. We leave the other two examples---as well as future releases of these and other products---as an exercise for the reader. 

The information in this section is based on publicly available documentation from the relevant vendors, not direct testing of the products. We list key points that currently available documentation leaves unclear, under the heading ``What We Still Don't Know'' at the end of each subsection. These lists are not comprehensive, but highlight notable points of ambiguity that hinder a complete evaluation of the E2EE compatibility of these AI integrations.

The latest versions of two of the products that we cover here were launched since the last version of this paper was published, less than three months ago. So, while our case-specific description and analysis will remain illustrative, some details may soon be out of date. Readers seeking to understand the implications of our work for the latest technologies should check recent product releases and may conduct their own analysis based on our generalizable analytical framework and recommendations.
\subsubsection{Apple Intelligence}
Apple introduced a new product called Apple Intelligence in late 2024, which provides new OS-integrated AI capabilities across the Apple ecosystem.\footnote{Apple Intelligence's initial release was in October 2024; the ChatGPT integration was launched in December 2024~\cite{applepressrelease, applepressrelease2-chatgpt}.} Apple Intelligence is able to process and perform AI tasks on E2EE content~\cite{applepressrelease}.
According to its online documentation, Apple's plan does not involve \emph{training} any models on any ``users' private personal data or user interactions,'' which presumably excludes E2EE content from use for training~\cite{appledeviceandservermodels}.

\paragraph{What data on Apple devices is E2EE?} Applications across the iOS ecosystem differ on the confidentiality guarantees they offer. Our work focuses only on data that is protected with E2EE, so in this section, we explain what data in Apple devices is E2EE. There are two broad classes of applications for Apple devices: native applications, which are managed by Apple and come built-in with the device; and third-party applications, which are downloaded from the App Store. A few native applications are E2EE by default, such as iMessage, Facetime, and Keychain. However, storage on iCloud is not E2EE by default, so most native applications that rely on cloud storage are not E2EE. For additional security, users have the option of enabling \emph{Advanced Data Protection (ADP)}~\cite{advanceddataprotection}, a feature that enables E2EE for the majority of iCloud data, such as photos, notes, voice memos, etc. Some iCloud data is not E2EE even when ADP is activated, such as mail, contacts, and calendars.\footnote{We refer readers to~\cite{advanceddataprotection} for a complete description of the confidentiality guarantees for various data types and native applications in iOS, both in the default case and when ADP is activated.} In this section, we refer to native applications for which ADP enables E2EE as \emph{ADP-E2EE applications}. In addition, ADP enables E2EE for certain Siri features, such as Siri shortcuts and other information (e.g., settings and personalization). It is not clear to us whether Siri transcripts are E2EE or not, even with ADP enabled.

The confidentiality guarantees of third-party applications\footnote{I.e., applications that run on Apple devices but are not developed by Apple.} on Apple devices vary by application. Some third-party applications are E2EE, such as Signal~\cite{signal}, and WhatsApp~\cite{whatsapp}.

\paragraph{What data can Siri and Apple Intelligence access?}
Siri and Apple Intelligence will be able to process users' content drawn from any application where the application developer has ``integrated'' Siri and Apple Intelligence functionalities~\cite{appleappintegration}, \emph{and} the user has not deactivated this integration in their settings~\cite{applesiriprivacy}. That is, if an application supports the integration of Siri and Apple Intelligence, then integration/access is on by default unless a user turns it off in their settings.

\label{sec:apple:usecases}
\paragraph{Possible use cases involving E2EE data}
We highlight two key contexts in which E2EE data is used for inference in Apple Intelligence. First, Apple Intelligence may be used to process user content produced inside an E2EE application. For example, long sequences of messages in iMessage chats may be processed by AI to produce a short ``summary'' (say, when a user has been away from a group chat, to help the user catch up)~\cite{applepressrelease}. Using Apple Intelligence or Siri inside any third-party E2EE or (if users have ADP enabled) any ADP-E2EE application could involve interaction with E2EE data. 

The second way in which Apple Intelligence may process E2EE data is through its content generation features, which may be accessible from within E2EE applications---this includes third-party E2EE and ADP-E2EE applications if ADP is enabled. For instance, image and emoji generation can be based on user-uploaded images and accessed through the iMessage app. The generated images and emojis may be based on E2EE content if ADP is enabled.

\paragraph{Three-Tiered Approach}
\label{sec:apple:techdetails}
As users interact with Apple applications on their iPhones, iPads, or Macs, these applications may submit queries to the Apple Intelligence system.
These queries, often a function of user data---and sometimes a function of user E2EE data---are then processed, and responses are sent back to the user (e.g., producing short summaries of long text or updating a contact's information).

Apple employs a three-tiered approach to handling user queries: (I) \emph{local models} which process simple queries entirely on-device; (II) a \emph{server model}, which processes more complex queries in a cloud-based trusted execution environment hosted by Apple; and (III) sending data to a \emph{third-party model}, OpenAI's ChatGPT, which processes outsourced queries that could benefit from ``ChatGPT’s broad world knowledge''~\cite{applepressrelease}. Each time a query is sent to ChatGPT, user consent is requested.\footnote{Our understanding is that permission prompts can be disabled with the exception of when files are sent to ChatGPT, in which case the user will always be asked \cite{openaisupport-sirichatgptext}. \label{fn:permission-prompts}}

These three cases are coordinated by a local \emph{orchestration layer}, which examines every query and determines how to process it: under Case I, II, or III. The orchestration layer prioritizes these cases in the order presented above: ``many of the models'' used to process queries are on-device~\cite{applepressrelease}~\cite{applepressrelease2-chatgpt}, only delegating ``more sophisticated requests'' to the more advanced model~\cite{pcc}, and ChatGPT is only summoned through Siri ``when helpful''~\cite{applepressrelease}. As we discuss in Section~\ref{sec:apple:what-we-still-dont-know}, Apple has not clarified which models process which types of queries, nor the criteria used by the orchestration layer to decide how a given query should be processed\footnote{We know that all images generated with Image Playground are created on device \cite{applepressrelease}.}. We now discuss what we do know about the three cases in more detail.

\begin{figure}
    \centering
    \includegraphics[width=0.9\linewidth]{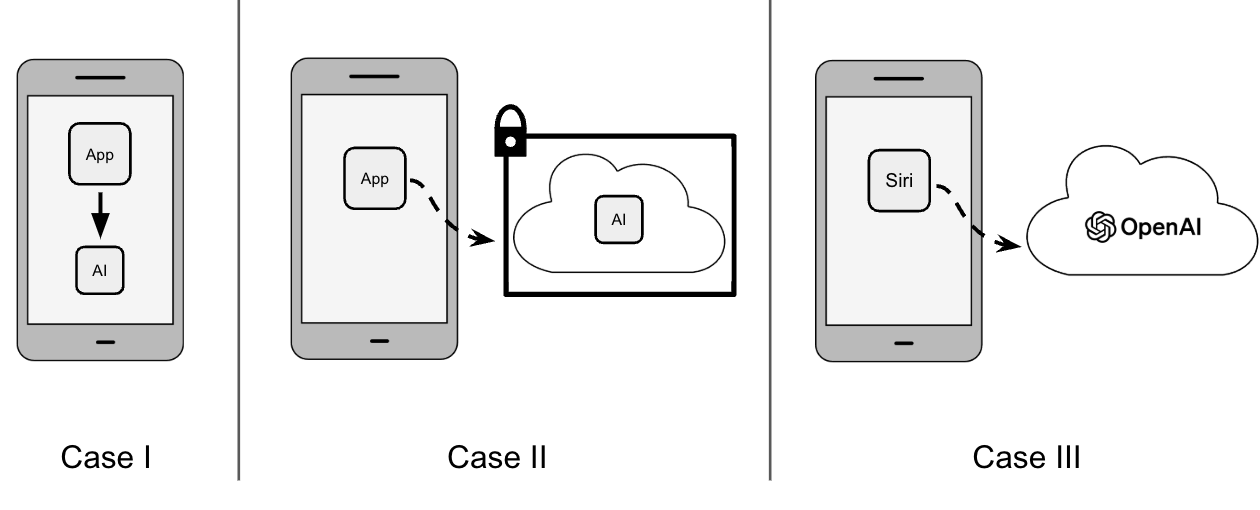}
    \caption{Apple's three-tiered approach}
    \label{fig:}
\end{figure}

\paragraph{Case I (Local model)} Apple Intelligence's first processing method consists of a series of small AI models stored entirely on a user's device~\cite{appledeviceandservermodels}. While some of these models are bespoke for certain tasks (e.g., code completion), their core local model consists of an approximately three-billion-parameter general-purpose language model, which can be adaptively fine-tuned for particular language tasks (e.g., summarizing text).

\paragraph{Case II (Server model)} Tasks that are ``sophisticated'' enough to fall in Case II are outsourced to a larger model, which is hosted on Apple's servers~\cite{pcc}. While the full details of this model have not been made public, we refer to~\cite{trailofbitsapplemodels} for an in-depth third-party analysis of possible architectures by Trail of Bits.
Outsourcing queries to a cloud-based model places---including E2EE data---undermines E2EE confidentiality. Apple offers some discussion of this heightened risk and proposes to use a trusted execution environment (TEE)\footnote{See Section~\ref{sec:background:trusted_hardware} for an introduction to TEEs.} to perform server-side inference more securely~\cite{pcc}.

Apple has built custom trusted execution environments in their data centers---which they call \emph{Private Cloud Compute (PCC)}---to host (and perform inference with) their server-side model. 
User devices communicate directly with Apple's PCC and supply plaintext inputs to the model, which are evaluated inside the PCC (i.e., inside Apple's TEE). 
Apple employs several additional measures to enhance the security of the PCC computation nodes; some of these measures appear to be designed partly in response to security considerations such as those outlined in Section~\ref{sec:background:trusted_hardware}.\footnote{For example: the PCC nodes leverage many of the hardware security technologies used by Apple's end-user products (such as their Secure Enclave Processor (SEP) and Secure Boot); the nodes are stripped of most functionality (except for security guardrails and inference capabilities) to minimize potential attack vectors; and user data (which is kept exclusively inside the PCC during computation) is deleted immediately once a response is returned to the user~\cite{pcc}. Apple additionally relies on a number of techniques to address supply-chain attacks and other backdoors, such as publishing every PCC production build in append-only, tamper-proof logs~\cite{pcc}, and X-ray imaging during assembly for anomaly detection~\cite{applepcc-hardwareintegrity}.}

\paragraph{Case III (Third-party model)} Finally, Apple Intelligence sends certain user queries to ChatGPT. This feature is turned off by default, but can be manually enabled in settings~\cite{applesupport-chatgptext}. When the feature is enabled, Siri examines each request and determines if ChatGPT ``might be helpful to get the information'' to respond to it~\cite{applesupport-siri}. The types of requests for which Siri may invoke ChatGPT may include direct prompts to ChatGPT, text generation, and summarization \cite{applesupport-siri}. If either ChatGPT is directly invoked or Siri deems ChatGPT could be ``helpful'' \cite{applepressrelease}, the user is asked in a pop-up prompt whether they would like to send the request to OpenAI for processing.\footnote{For an example of what this looks like, see \href{https://www.apple.com/newsroom/images/2024/06/introducing-apple-intelligence-for-iphone-ipad-and-mac/article/Apple-WWDC24-Apple-Intelligence-Siri-ChatGPT-request-240610_inline.jpg.medium_2x.jpg}{this} example from the Apple Intelligence press release \cite{applepressrelease}.} By default, this prompt occurs every time a request could be sent to ChatGPT, but can be disabled in settings\footnote{See footnote~\ref{fn:permission-prompts}.} \cite{applesupport-siri}.
Apple collects limited metadata about the queries sent to ChatGPT, including the number and size of requests~\cite{applesiriprivacy}.

Upon enabling the ChatGPT extension in Siri, users may choose to continue with or without logging in to their OpenAI account.\footnote{For an example of what the log-in prompt looks like, see~\cite{openaisupport-sirichatgptext}.
}
If the user doesn't log in, query processing is consistent with Apple Intelligence's privacy policies: OpenAI is required to process queries ephemerally and must not use data from requests to train its models~\cite{applesupport-chatgptext}.\footnote{When a conversation is continued in the ChatGPT app or website, queries are processed under OpenAI's privacy policy, regardless of whether the user was initially signed in through Apple Intelligence or not~\cite{openaisupport-appleintelligence-datahandling}.} However, without logging into their OpenAI account, the user would not have access to any premium features linked to their OpenAI account~\cite{openaisupport-appleintelligence-datahandling}. If, on the other hand, the user chooses to log in to their OpenAI account, query processing is governed by ChatGPT's privacy policies and the user's OpenAI account settings~\cite{applesupport-chatgptext}---not Apple's policies. As such, for logged-in users, OpenAI may log requests and use query data to train its models if the user's OpenAI account permits such processing (even if this is inconsistent with Apple's policies)~\cite{applesupport-chatgptext}.\footnote{ChatGPT accounts have model training on by default~\cite{openai-datacontrols}.}

\paragraph{What We Still Don't Know}
\label{sec:apple:what-we-still-dont-know}
Based on the available documentation at the time of writing, significant details remain unclear about Apple's system. Open questions, to which the answers could be essential to independently evaluate the system's security risks and compatibility with E2EE, include the following.
\begin{newitemize}
    \item Will users be aware each time Apple Intelligence is invoked?
    \item How is it determined whether any given query will fall under Case I, II, or III?
    \item Will users be aware of how each of their queries are processed---under Case I, II, or III?\footnote{We know that users will be prompted before sending any query to ChatGPT, but the delegation protocol between on and off-device processing remains unclear.}
    \item How will queries be processed for users who opt into Advanced Data Protection? There does not appear to be documentation regarding whether the processing of queries from these services (when Advanced Data Protection is enabled) will be handled differently by Apple Intelligence.
\end{newitemize}

After our analysis and recommendations, we will return to the Apple system in Section~\ref{sec:recommendations:apple-intelligence}.

\subsubsection{Samsung's Galaxy AI}
Samsung introduced a new series of AI-integrated phones in January 2025~\cite{samsung-s25pressrelease}. Several new phones, such as the Galaxy S25 Ultra, Galaxy S25+, and Galaxy S25, will come with integrated AI assistants~\cite{samsung-s25pressrelease}: Samsung's own Galaxy AI, some of whose features leverage Google's Gemini~\cite{samsung-s25pressrelease}. While some previous Samsung phones were able to tap into certain Galaxy AI features, these newer models provide significant additional use cases.

\paragraph{What data on Samsung devices is E2EE?}
E2EE applications on Samsung phones are either preinstalled on the device or can be downloaded from the Galaxy Store. Google Messages, for instance, comes with the phone and is E2EE by default between Google Messages users~\cite{samsung-defaultmessaging}~\cite{googlemessages-e2ee}. Cloud storage on Samsung devices is not E2EE by default. However, users have the option of enabling \textit{Enhanced Data Protection} (EDP), a feature which enables E2EE for call log, message, clock, settings, and app backups~\cite{samsung-enhanceddataprotection}.~\footnote{Large amounts of data such as message attachments may not be backed up under E2EE, even with Enhanced Data Protection~\cite{samsung-enhanceddataprotection}.}

\paragraph{Possible use cases involving E2EE data} Galaxy AI can process content in E2EE applications, including via the Gemini integration which can perform tasks across Google, Samsung, and third-party apps~\cite{samsung-s25pressrelease,samsung-galaxyaifaqs}. If users have Enhanced Data Protection enabled, Galaxy AI may also process additional content that is E2EE under EDP for purposes such as generating call transcripts and others~\cite{samsung-s25pressrelease}.  

\paragraph{Two-Tiered Approach}
Galaxy AI queries are either processed on-device or outsourced to a cloud model. Samsung's documentation notes that Google Messages features, ``most translations'', and ambient wallpapers are processed on-device, whereas most other features, including auto-summarizing and image editing, are cloud-based~\cite{samsung-galaxyaifaqs}.

On-device AI processes data locally with the Personal Data Engine, which is secured by hardware measures such as Knox Vault~\cite{samsung-knoxvault}. When processing on-device, queries and data are not used for training~\cite{samsung-dataprocessing-galaxyai}.

According to Galaxy AI documentation as of January 2025, cloud models can use user query data for training~\cite{samsung-dataprocessing-galaxyai}. However, as explained at Samsung's annual Galaxy Unpacked event in the same month, when Galaxy S25 series phones process queries in the cloud, user data is only used to respond to the query, is deleted as soon as the query has been responded to, and will not be used for training~\cite{samsung-galaxyunpacked}.
Samsung partnered with Google Cloud in 2024 to provide Gemini-powered AI tools~\cite{samsung-googlecloud-partnership}. Several aspects of Galaxy AI's cloud processing arrangement and security measures remain unclear, as further discussed below.

Galaxy AI is on by default on these newer phones, but some features require users to log in to their Samsung or Google accounts~\cite{samsung-galaxyaifaqs}. For example, Cross App Action (a feature that can perform tasks across different apps, including Google, Galaxy native, and third-party apps such as WhatsApp and Spotify) will require users to log into their Google account~\cite{samsung-aifeatures}.\footnote{The Now Brief and Now Bar features, which provide users with relevant updates, will require users to log into their Samsung accounts~\cite{samsung-aifeatures}. And other features such as the Audio Eraser and AI Select also require a Samsung account~\cite{samsung-aifeatures}.} Users have the options to disable cloud-based features while maintaining on-device processing capabilities, and to disable all AI features, in their settings~\cite{samsung-galaxyaifaqs}.

\paragraph{What We Still Don't Know}
\begin{itemize}
    \item Does all Galaxy AI cloud processing take place on Google Cloud or just Gemini-based features?
    \item What security measures are used for Galaxy AI's cloud processing? How are encryption and trusted hardware employed, if at all?
    \item What notices and disclosures are given to the user?
    \item Are users aware of when AI is invoked?
    \item Are users aware of how AI queries are being processed—locally or on the cloud?\footnote{Users are able to select if they want to turn off AI features or turn off cloud-based AI features in their settings~\cite{samsung-galaxyaifaqs}, but it remains unclear if users are told upfront how their queries are being processed.}
\end{itemize}

\subsubsection{Meta AI in WhatsApp}
Meta features an integrated AI assistant in the E2EE messaging platform WhatsApp~\cite{whatsapp-faqmetaai}. This case, software integration instead of device integration, is slightly different than the previous two. 

Meta AI can either be accessed in a standalone chat (between a user and the AI assistant) or it can join other conversations when it is summoned explicitly by a user sending a message containing the term ``@MetaAI''~\cite{whatsapp-faqmetaai}. In the latter case, Meta AI is only able to read messages that mention @MetaAI or explicit replies to these messages~\cite{whatsapp-faqmetaai}. These boundaries are noted in the initial disclaimer\footnote{See screenshots in Appendix~\ref{sec:metaai-disclaimers}.} the first time Meta AI is invoked, but do not appear afterward. Users are also given the option to delete AI messages by typing ``/reset-all-ais''~\cite{whatsapp-faqusingmetaAI}, though as noted below, it is not clear what exactly this does.

Meta AI always involves cloud processing; it does not run locally on user devices. Meta AI's cloud processing infrastructure is not clear to us, although Meta has noted that the operation of Llama---on which Meta AI is based---is supported by ``roster of partners ... across the hardware and software ecosystem'' including a number of third-party cloud providers~\cite{metaai-cloudmodels}. It is trained on a combination of publicly available online information, licensed information, and information shared on Meta products and services~\cite{metaai-faqs}. Queries sent from WhatsApp to Meta AI may be used for training, and this training data is kept by Meta ``as long as [they] need it on a case-by-case basis''~\cite{metaai-faqs}; there is no option for users to opt out of this usage.

\paragraph{What We Still Don't Know}
\begin{itemize}
    \item Where does Meta AI's cloud processing take place? How are Meta's ``partners ... across the hardware and software ecosystem'' involved, if at all?
    \item What security measures, such as encryption and trusted hardware, are used for cloud processing?
    \item What exactly happens when users ``reset'' and delete chats with AIs?
\end{itemize}

\section{Our Definitions and Taxonomy}
\label{sec:defs}

\subsection{Terminology}
\label{sec:defs:e2ee-terms}

Next, we define nine key terms we use in our recommendations and analysis. 
In addition to the definitions in this section, we rely throughout on key definitions introduced in Section~\ref{sec:background} (Background), such as \emph{encryption}, \emph{plaintext}, \emph{ciphertext}, \emph{model}, \emph{inference}, \emph{training}, and others.
While Section~\ref{sec:background} defines basic terms from established literatures, the present section focuses on clarifying the terminology essential to make our analysis and recommendations precise within the scope of this paper. As such, our claim is not that the following are the universal definitions of these terms, but that these are the meanings we use in this paper.

\paragraph{Endpoint} An \emph{end} is either the sender or a recipient of a message. %
An \emph{endpoint} encompasses an ``end'' user and the user's device or devices (that the user has authenticated with a client application to send and receive messages\footnote{Endpoint identification is an important security consideration in practice, which is largely out of scope for this paper~\cite{hale2022end}.}).

\paragraph{Endpoint-local} We say that an operation, computation, or process is \emph{endpoint-local} if it takes place entirely on the user's physical device. Endpoint-local computation is often done with a client application. This definition excludes any forms of implicit or virtual locality (e.g., a user-specific private cloud container) as well as any transmission of data off the user's physical device.

\paragraph{Primary application content} For a given application, the \emph{primary application content} is the content within the application that is part of the main service(s) that the application provides. 

\begin{quote}
    For example, in a messaging app, the primary application content consists of communication contents (including text, images, voice messages, and any other content types supported by the app). In a videoconferencing app (like Zoom), the primary application content consists of the contents of video calls (including audio and video but also other in-call content such as polls or screen-sharing). %
\end{quote}

\paragraph{E2EE application} We use \emph{E2EE application} broadly to mean any application that makes claims that it offers E2EE in any part of the application functionality---for primary application content or otherwise.

\paragraph{Metadata} Information that is auxiliary to the content offered in an application is called \emph{metadata}.\footnote{The distinction between content and metadata is complex and context-specific  \cite{bellovin2016metadata}.} In E2EE applications, metadata is generally not encrypted.  

\begin{quote}
    For a messaging service, this includes who is communicating with whom, at what times and how frequently messages are exchanged, message lengths, and potentially many other types of data associated with a message---but not the message content.   
\end{quote}

\paragraph{E2EE content} We write \emph{E2EE content} or \emph{E2EE data} to refer to any data which a service provider explicitly or implicitly indicates is protected by E2EE in an application, and any derivatives of such data.\footnote{Derivatives include any mathematically correlated function of such data (excluding bit-length).} Many claims of E2EE by service providers are not very specific regarding which data are E2EE---in such cases, we treat a generalized claim of E2EE in an application as an implicit claim that the primary application content is end-to-end encrypted.  We do not treat non-primary application content as E2EE content unless a provider has explicitly claimed that that type of content is E2EE. 

\begin{quote}
    For example, if a messaging provider claims that its app is E2EE, then the text, audio, images, videos, and any other message content exchanged by users---and any derivatives thereof---are E2EE content. In contrast, messaging metadata would not be E2EE content in this scenario.)
\end{quote}

\paragraph{Shared model} A shared model is an AI model that may be be used for inference by multiple users, and/or may be trained on multiple users' queries. In a given E2EE context, we describe a model as \emph{shared} if its users may include people and/or AI models other than the sender or the recipients of E2EE content.

\paragraph{Per-user model} A per-user model is an AI model that can be used for inference exclusively by a single user, and that is continuously improved with queries from only that user. Note that the latter implies that data from a particular user is not used to train the per-user models of other users. As further discussed in Section~\ref{sec:technical-implications:private-inference}, this definition encompasses both ``physical'' and ``logical'' implementations, such as completely separate models for each user or a single foundational model dynamically refined during inference, respectively.

\paragraph{Third party} With respect to a particular piece of E2EE content, a \emph{third party} is defined as any party which is neither the sender nor intended recipient of that content.

\begin{quote}
For example, if Alice is sending a message to Bob using a messaging platform $A$, then $A$ is a third party with respect to that message. If the platform $A$ outsources some processing to another company $B$, then $B$ is also a third party to users' conversations.

If Alice is querying an AI assistant Hal directly, (e.g., ``Hal, what will the weather be like this evening?'') and Hal is a cloud-based AI assistant offered by platform $A$, then we do \emph{not} consider $A$ (and Hal) to be a third party with respect to that query.

\end{quote}

\subsection{A Taxonomy From ``Strict E2EE'' to ``No E2EE''}
\label{sec:buckets}

In this section, we present a taxonomy of applications ranging from ``Strict E2EE'' to ``No E2EE,'' consisting of five categories in decreasing order of the degree to which they maintain the integrity of E2EE. As discussed in Section~\ref{sec:background:e2ee}, modern E2EE applications typically involve additional features and complexity beyond those explicitly modeled by formal definitions. As such, they cover the full range of our first four categories.

\medskip

\textbf{Category 1 (Strict E2EE)} refers to applications where \emph{no E2EE content is seen or used by any third party at any point in the functioning of the application, for all available configurations of application settings and features}.

\medskip
No major E2EE application meets Category 1's most stringent level of confidentiality, at the time of writing. 
Certain commonplace features that enhance user experience involve the disclosure of (often small amounts of) E2EE content to third parties.\footnote{One example, already discussed in Section~\ref{sec:background:e2ee:usability}, is \emph{link previews}.} 
As such, even with these features, it has become common to refer to and advertise the application as ``end-to-end encrypted,'' both in industry and in academia, including among encryption experts. 
We refer to such features, which involve third-party processing of strictly limited types or quantities of plaintext-dependent content to provide a user-experience-enhancing feature, and which are commonplace in applications that provide the strongest E2EE guarantees in current practice, as \bi{caveat features}.\footnote{We recognize that term ``strictly limited'' is ambiguous here; we do not think it is realistic or helpful to articulate a precise threshold amount of processing data that would be acceptable. That said, we would \emph{not} consider a feature involving third-party processing of a majority of E2EE content to be ``strictly limited'' processing. We \emph{would} consider the caveat features that are commonplace in the most prominent E2EE applications in current practice to involve third-party processing that is ``strictly limited'' in the sense we mean here.}

\medskip

\textbf{Category 2 (E2EE in current practice)} encompasses a range of applications that do not qualify as ``Strict E2EE'' due to caveat features, but otherwise provide strong E2EE guarantees. 
To be in Category 2, \emph{no E2EE content may be seen or used by any third party at any point in the functioning of the application, \emph{except} for caveat features}, and applications must offer each user the option of disabling all caveat features in their own use of the application. (In other words, there must exist a configuration of application settings and features in which the application operates like a Category 1 application.)
We place Signal\footnote{Showing the preview image of a link shared in Signal violates strict E2EE, for example\cite{signal2019link}.} and WhatsApp\footnote{WhatsApp's (non-E2EE) backup feature violates strict E2EE, for example\cite{whatsappbackup}.} in Category 2. 

\medskip

The current and prevailing use of the term ``E2EE'' generally refers to Categories 1 and 2, both in industry and in academia, including among encryption experts.

\medskip

\textbf{Category 3 (E2EE with opt out)} refers to applications which offer strong E2EE features as an option, but either these features are off by default, or the application allows users to opt out of these features. Note, in contrast, that Categories 1 and 2 definitionally preclude E2EE being off by default and/or users turning E2EE off. A Category 3 application's E2EE features, \emph{when enabled}, must be strong enough that \emph{no E2EE content may be seen or used by any third party at any point in the functioning of the application, except for caveat features}. (That is, again, there must exist a configuration of settings and features in which the application operates like a Category 1 application.) We place Zoom~\cite{zoom} in Category 3, as it offers E2EE videoconferencing only when turned on in user settings. We consider Zoom to be in Category 3 since a user can set E2EE as their personal default in their account settings (which means calls that that user sets up will be E2EE unless the user explicitly indicates otherwise); what we focus on is whether E2EE is off by default \emph{for any given user account}.
Notably, Category 3 applications lack the important \emph{systemic protections} discussed in Section~\ref{sec:background:e2ee:systemic-e2ee-properties}.

\textbf{Category 4 (E2EE options available)} encompasses applications that offer E2EE as an option for limited features, but does not offer any configuration where all primary application content is protected by E2EE. That is, Category 4 applications are not E2EE because they \emph{always involve processing of primary application content by third parties in the functioning of the application} unless data is protected by E2EE as a narrow exception. Telegram allows users to turn on E2EE on a per-chat basis and it is always off by default~\cite{telegram}.

Finally, \textbf{Category 5 (No E2EE)} describes applications that fall in none of the above categories. Category 5 does not preclude the use of encryption: for example, applications that use \emph{encryption} but not \emph{end-to-end} encryption fall in Category 5. As later sections elaborate, Category 5 applications are outside the scope of this paper---unless the service providers make claims that the application does offer E2EE.

\section{Security Implications of Integrating AI with E2EE}
\label{sec:technical-implications}

Building an AI assistant into an E2EE application involves several layers of configuration which affect the ways in which content may be used, processed, and stored. 
Section~\ref{sec:technical-implications:ai} discusses how integrating AI assistants into E2EE applications raises significant security concerns due to the mechanisms by which AI assistants use and process data. Section~\ref{sec:technical-implications:private-inference} provides an overview of existing cryptography and privacy technologies that may (seem to) mitigate some of the aforementioned security concerns, and explain how these technologies are \emph{insufficient} to provide E2EE confidentiality for AI assistants processing E2EE content.

\subsection{Security Considerations for AI assistants}
\label{sec:technical-implications:ai}
As explained in Section~\ref{sec:background}, AI assistants interact with data during both training and inference. For an implementation of an AI assistant to be compatible with E2EE, it must uphold the guarantees of E2EE during both training and inference.

\subsubsection{AI Assistants Interacting with E2EE Data}
\label{sec:technical-implications:ai:attacks}

As discussed in Section~\ref{sec:background:ai}, a model's training data consists of vast amounts of information which can be continuously supplemented by data collected from users (e.g., queries, usage patterns) after the model is deployed and used for inference. This data would necessarily be plaintext as a model cannot be trained on encrypted text. As such, if user data is collected and incorporated into the training process, the model is inherently, itself, a derivative of E2EE content.

Additionally, models have been observed to reproduce portions of the training data in their outputs, at times producing significant portions of training data, sometimes verbatim or nearly verbatim  \cite{cooper2024files, petroni2019languagemodelsknowledgebases, mccoy2023much}. %
The reasons for this phenomenon are not fully understood, but it is believed by many experts to be inherent in current machine learning and AI development techniques. 
Training on E2EE content thus raises the security threat of models exposing sensitive information from the training data even during regular (non-adversarial) use, constituting leaks of user data.

The memorization and reproduction of training data in models also introduces the risk of individuals exploiting this behavior to extract certain data points and patterns. Prior work has shown that adversarial queries can expose information in the training data, which would directly compromise the confidentiality of E2EE data \cite{nasr2023scalable}. Individuals may potentially perform \emph{adversarial attacks}, involving using the model (often in a black-box manner) as a means to acquire private data. Adversarial attacks are an active research area, and come in various forms, including \emph{membership inference attacks}, where an adversary can determine whether a particular data point is part of the model's training set \cite{shokri2017membership, hu2022membership}; and \emph{training data extraction attacks}, where an adversary can recover elements of the model's training data often by generating candidate data points and using membership inference to verify whether the items are in the training set \cite{274574}.

\subsubsection{Key Considerations}

Given that integrated AI assistants may process potentially significant volumes of E2EE data, they are \emph{not} an example of caveat features (as defined in Section~\ref{sec:buckets}). How integrated AI assistants are implemented in an E2EE application directly impacts how strongly an application preserves E2EE, and the addition of integrated AI may demote an application's categorization within our taxonomy (Section~\ref{sec:buckets}). The risks outlined above underscore the following critical considerations for the implementation of AI assistants.
\begin{enumerate}
    \item \emph{Which users can query the models?} 
    To preserve the guarantees of E2EE, user messages should only be used to train models that can be queried by the senders and intended recipients of the messages.
    \item \emph{Which parties have access to the inputs and outputs from the model?} Intermediaries that process plaintext user queries before relaying them to the model provider may have access to inputs and outputs from the model, in the form of chat logs. To preserve the guarantees of E2EE, inputs and outputs that depend on user messages should only be visible to the intended recipients of the messages.
    \item \emph{Are the inputs to the model used for anything else besides inference?} Even if the no parties and no other users can query or access logs of the model, inputs to the model may be used for other purposes---for example, the model provider may collect aggregate statistics on user queries. To be compatible with E2EE, queries should be used exclusively to fulfill users' requests. In other words, inference should be a \emph{stateless} process.
\end{enumerate}

\subsection{How Privacy Technologies Can and Cannot Help}
\label{sec:technical-implications:private-inference}
The privacy and confidentiality concerns outlined in the preceding subsection are by now well known to the machine learning, privacy, and security communities, and there exist various techniques to enhance privacy (according to various metrics and definitions) during inference and training. 

Next, we evaluate whether the types of privacy enhancements that existing techniques offer meet E2EE confidentiality, considering a range of possible implementations. We find that existing privacy-enhancing techniques \emph{cannot} address the security concerns we raise around integrating AI assistants with E2EE applications. We consider the following privacy-enhancing techniques: \emph{endpoint-local training and inference}, \emph{fully homomorphic inference (FHE)}, \emph{trusted execution environments (TEEs)}, \emph{per-user models}, \emph{data sanitization}, \emph{differential privacy}, \emph{federated learning}, and \emph{multi-party computation (MPC)}. While some of the techniques and implementations that we consider reflect existing practices, others represent novel configurations.

\begin{figure}
    \centering
    \includegraphics[width=0.9\linewidth]{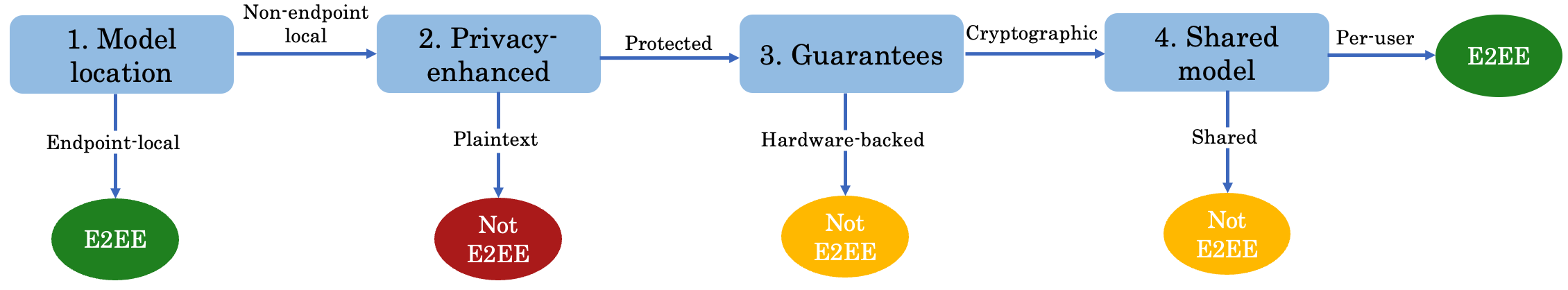}
    \caption{Evaluation framework for implementations of AI assistants }
    \label{fig:framework-diagram}
\end{figure}

Underlying our evaluation is a generalizable analytical framework designed to \emph{identify the key design considerations for implementing AI assistants in E2EE environments}, structured around four key considerations which are detailed below.
Our framework can be used to evaluate and compare different implementations of AI assistants with (or without) diverse privacy technologies, and the degree to which they uphold E2EE security, generalizing beyond the specific technologies that we examine in this paper. The four key framework considerations are interdependent: at each consideration point, design choices can either affirm the configuration's compliance with E2EE or breach it (thereby rendering further considerations irrelevant). Figure~\ref{fig:framework-diagram} summarizes our evaluation framework as a flowchart, and Table~\ref{tbl:technical-configs-framework-examples} summarizes our analysis of various AI assistant implementations' compatibility with E2EE under our framework. In Appendix \ref{sec:e2ee-features-eval}, we present an extended framework for evaluating the compatibility of arbitrary features (as opposed to just AI assistants) with E2EE, which generalizes the framework from this section and encompasses three levels: technical implementation, data inputs, and user control. %

\paragraph{Consideration \#1: Where is the model located?}
The first design parameter is the location where data is processed, i.e., where the model is hosted.
We distinguish between \emph{endpoint-local} and \emph{non-endpoint-local} designs, using the definitions from Section~\ref{sec:defs:e2ee-terms}. 

Endpoint-local models are fully compatible with E2EE, and are the safest way to perform inference and training, since no messages are ever handled by an external server. (This holds true provided that E2EE messages are only used to train endpoint-local models of the sender and recipients, and not used to fine-tune endpoint-local models in other devices.) A further advantage of endpoint-local models is that they allow the use of unmodified plaintext messages when training, and thus can train upon full-quality data.

Using endpoint-local models is not simple, however, as most consumer grade devices are unable to support the hefty computations involved in running an AI assistant.
Local processing is feasible in limited circumstances: e.g., in certain use cases of Apple Intelligence and Samsung's Galaxy AI (see Section~\ref{sec:apple}). 
But even with custom hardware, not all AI assistant queries can be supported with the limited computational faculties of a smartphone, and many features require sending plaintext messages off the device. 

When non-endpoint-local models are used to answer queries, the compatibility with E2EE depends on Considerations \#2 and \#3 below.

\paragraph{Consideration \#2: If the model is non-endpoint-local, is its processing of user queries privacy-enhanced, or in plaintext?}
For non-endpoint-local models, the confidentiality of data depends on whether processing of queries occurs in \emph{plaintext} or in some \emph{privacy-enhanced} form (i.e., encrypted or contained in secure hardware) on the servers.
Plaintext inference and training is not compatible with E2EE, as they would expose user data to third parties.
Privacy-enhanced processing provides additional protection. 
There are various technologies that could implement some kind of ``privacy enhancement''---such as fully homomorphic encryption (FHE), and trusted execution environments (TEEs)---which differ on their confidentiality guarantees, performance, and other practical considerations. The compatibility of privacy-enhanced processing with E2EE depends on the nature of the confidentiality offered by the technology being used, which is the focus of Consideration \#3.

\paragraph{Consideration \#3: If the model is non-endpoint-local and performs privacy-enhanced processing, what is the type of the confidentiality guarantees?}
As discussed in Section~\ref{sec:background:trusted_hardware:tees_v_e2ee}, different security technologies---such as TEEs and encryption---provide different types of confidentiality under different conditions.
Established \emph{cryptographic} techniques tend to provide the same type of security as E2EE under types of conditions, and are therefore more likely to be compatible with E2EE.

Fully homomorphic encryption (FHE) provides an instructive example. FHE enables arbitrary computations to be performed on encrypted data without the need to decrypt the data, and yields the result of the computation still in encrypted form. FHE makes it possible to ``outsource'' a computation on private data without revealing anything about the data to the outsourcing provider. FHE would thus allow non-endpoint-local models to process user messages for inference without decrypting them---at least in theory.

FHE schemes are far less efficient than typical encryption schemes like those used in E2EE applications. Consequently, computationally intensive operations like inference are prohibitively slow under FHE, and would not result in practical turnaround times for users.
For example, Zama, a company working on applications of FHE in blockchain and AI, estimates that, at present, running an ``average-sized LLM'' under FHE costs about \$5,000 per word
\cite{zama_e2ee_llm_projections}. While FHE is not a presently viable solution, it is an active area of research; substantial breakthroughs in the state of the art in the machine learning, hardware, and cryptography fronts will be necessary to eventually make this approach practical.

\emph{Hardware-backed} security is the second major class of confidentiality guarantees, which involve using trusted hardware, such as TEEs (Section~\ref{sec:background:trusted_hardware}), to process user data. In this case, the AI model would be hosted inside a non-endpoint-local TEE. %
Processing inside the TEE would happen in plaintext, while still being shielded from outside entities by the hardware protections of the TEE.

As described in Section~\ref{sec:background:trusted_hardware:tees_v_e2ee}, 
hardware-backed security is different from E2EE security.
Hardware solutions represent a substantial security/privacy improvement over plaintext inference,
which may well be a fruitful avenue for exploration, beyond the scope of this paper. Our scope is to stress that the importance of clearly distinguishing them from E2EE and cryptographic security guarantees.

\paragraph{Consideration \#4: If the model is non-endpoint local, is it a shared or per-user model?}
Any model training introduces an additional critical consideration: who can query models trained with whose E2EE data? Since a model is a derivative of its training data, it is important that queries over this training data are consistent with the guarantees of E2EE.

There are two possible configurations for this design choice: \emph{shared} or \emph{per-user} models, as defined in Section~\ref{sec:defs:e2ee-terms}. 
Shared models are not compatible with E2EE, since users would be able to receive inference outputs that are a function of other users' E2EE data, which violates E2EE confidentiality. In practice, this means that E2EE data may be exposed to entities that can query the model (e.g., due to the possibility of leakage of training data and adversarial attacks). As we discuss in more detail in Section~\ref{subsubsec:on-privacy-preserving-training}, mitigations for adversarial attacks (e.g., differential privacy) are insufficient to fully address this concern. Shared models are incompatible with training on E2EE content, irrespective of the implementation.

On the other hand, per-user models can be compatible with E2EE, provided that applications enforce strict isolation between user data, and that any components that depend on E2EE data must themselves be adequately privacy-enhanced in the sense outlined in Considerations \#2 and \#3. There are a number of ways an application could support per-user models. For example, a service could maintain a separate copy of a foundation model for each user in the system, which is fine-tuned with that user's data. This approach would maintain strict isolation, but require large amounts of resources for applications with many users.
A more practical approach is to use \emph{retrieval augmentation} techniques: the application hosts a single foundation model, which is not trained on user data, but is dynamically refined during inference using user-dependent data sources~\cite{martineau_what_2023}.\footnote{See~\cite{wutschitz2023rethinking} for more on retrieval augmentation.} The application could store E2EE data for every user locally (e.g.,~\cite{martineau_what_2023}), and retrieve and incorporate it locally during inference---temporarily fine-tuning the model. After the query is processed, the base model should revert to its prior state, which was not dependent on any E2EE data, resulting in a \emph{stateless} augmentation process. While this reversion or statelessness would largely address the specific concern of Consideration \#4---namely, which users can query the models trained with the E2EE data of which users---the storage and processing of the per-user E2EE data and the temporarily fine-tuned model (which is itself E2EE data) would, in turn, need to be scrutinized under Considerations \#1--\#3 and comply with the recommendations we put forward throughout this paper.

\paragraph{Applying this framework} A specific implementation of an AI assistant can be evaluated with respect to each of the four design choices in our framework. From the results of this evaluation, we can determine the implementation's compatibility with E2EE. For example, the analysis of local inference would be as follows: 

\begin{quote}
    (1) local, (2) N/A, (3) N/A, (4) N/A, therefore fully compatible with E2EE. 
\end{quote}

Example analyses of a range of possible implementations are presented in Table~\ref{tbl:technical-configs-framework-examples}.

\definecolor{carnelian}{HTML}{B31B1B}
\definecolor{forestgreen}{HTML}{228B22}
\definecolor{yellow}{HTML}{FFC000}

\begin{figure*}[ht!]
    \centering
    \footnotesize
    \renewcommand{\arraystretch}{1.1}
    {
      \begin{NiceTabular}{l|llll}
      \toprule
      \textbf{Implementation} & \textbf{Location} & \textbf{Privacy-enhanced} & \textbf{Guarantees} & \textbf{\makecell{Shared or \\ per-user}}\\
      \toprule
      \makecell[tl]{On-device, fine-tuned model} & Endpoint local & No & N/A & N/A\\
     \midrule
     \makecell[tl]{Server-side plaintext inference \\ without fine-tuning} & Non-endpoint local & No & N/A & N/A\\
     \midrule
     \makecell[tl]{Server-side inference using \\ FHE without fine-tuning} & Non-endpoint local & Yes & Cryptographic & N/A\\
     \midrule
     \makecell[tl]{Server-side inference on separate \\ per-user fine-tuned models \\ on TEEs} & Non-endpoint local & Yes & Hardware-backed & Per-user \\
     \midrule
     \makecell[tl]{Server-side inference using \\ FHE on a global model \\ fine-tuned with DP} & Non-endpoint local & Yes & Cryptographic & Shared \\
     \midrule
     \makecell[tl]{Server-side inference using \\ FHE on a global model} & Non-endpoint local & Yes & Cryptographic & Shared \\
     \midrule
      Apple's proposal & Non-endpoint local & Yes & Hardware-backed & N/A \\
     \bottomrule
     \CodeAfter
     \tikz \fill [forestgreen] (2-|1) rectangle ($(3-|1)!0.03!(3-|2)$) ;
     \tikz \fill [carnelian] (3-|1) rectangle ($(4-|1)!0.03!(4-|2)$) ;
     \tikz \fill [forestgreen] (4-|1) rectangle ($(5-|1)!0.03!(5-|2)$) ;
     \tikz \fill [yellow] (5-|1) rectangle ($(6-|1)!0.03!(6-|2)$) ;
     \tikz \fill [yellow] (6-|1) rectangle ($(7-|1)!0.03!(7-|2)$) ;
     \tikz \fill [carnelian] (7-|1) rectangle ($(8-|1)!0.03!(8-|2)$) ;
     \tikz \fill [yellow] (8-|1) rectangle ($(9-|1)!0.03!(9-|2)$) ;
\end{NiceTabular}}
\captionof{table}{Example analyses of different implementations of AI assistants under our framework. \textcolor{forestgreen}{Green} tags denote an implementation compatible with E2EE, \textcolor{yellow}{yellow} tags denote an implementation that is not compatible with E2EE but has additional protections, and \textcolor{carnelian}{red} tags denote implementations that are most vulnerable.}
\vspace{-0.3cm}
\label{tbl:technical-configs-framework-examples}
\end{figure*}

\paragraph{Connection to our taxonomy}
Recall that Section \ref{sec:buckets} describes a taxonomy from ``Strict E2EE'' to ``No E2EE.''
The (in)compatibility of an AI assistant implementation with E2EE, as assessed under the evaluation framework just presented, directly impacts an application's categorization under our taxonomy. An application's original features, before integrating an AI assistant, determine what we call its \emph{initial} categorization; if AI features are added, a fresh evaluation and categorization are necessary, and the new categorization will depend on the implementation details highlighted in our evaluation framework.

The addition of a non-E2EE-compatible AI assistant implementation to an application that is initially in Category 1 or 2 will result in the application being ``demoted'' out of those two categories, as the added AI feature will render E2EE content no longer fully protected in the way that E2EE promises.
In more detail:
\definecolor{carnelian}{HTML}{B31B1B}
\begin{itemize}
    \item If the AI assistant implementation is fully compatible with E2EE (\textcolor{ForestGreen}{green} tags in Table~\ref{tbl:technical-configs-framework-examples}), the application maintains its initial category (from before the AI feature was added).
    \item If the AI assistant implementation is \emph{not} fully compatible with E2EE (\textcolor{yellow}{yellow} or \textcolor{carnelian}{red}  tags in Table~\ref{tbl:technical-configs-framework-examples}), the application is demoted to Category 3, 4, or 5, as follows: If the AI assistant is \emph{off by default} or \emph{provides an option to turn the AI assistant off application-wide}, the application is demoted to Category 3. If the AI assistant is \emph{on by default} and only \emph{provides an option to turn off in specific contexts} (e.g., for individual chats), the application is demoted to Category 4. If the AI assistant \emph{cannot be turned off}, the application is demoted to Category 5.
\end{itemize}

\subsubsection{Discussion on Privacy-Preserving Training}\label{subsubsec:on-privacy-preserving-training}

This section considers a range of privacy-preserving training methods---differential privacy, data sanitization, federated learning, and multi-party computation---and discusses how the types of privacy guarantees they provide are different from E2EE confidentiality. None of these techniques can guarantee E2EE confidentiality for training shared models on E2EE content.

\textbf{Differential privacy} (DP) is a method for designing algorithms to compute statistics over datasets such that the results do not depend ``too much'' on any individual data entry. 
In contrast with encryption's binary security guarantee, DP is parameterized by an ``acceptable'' amount of ``privacy loss,'' representing dependence upon individual data entries
(encoded via a parameter $\varepsilon$)
~\cite{DBLP:conf/tamc/Dwork08}.
Applying DP to machine learning is an active area of research; training with DP incurs a computational overhead and loss in accuracy compared to a non-DP model~\cite{nist_dp_ml,Ponomareva_2023}.
Though DP is often discussed alongside cryptographic techniques, DP offers a categorically different security guarantee than encryption, applicable under different conditions.

\textbf{Data sanitization} aims to remove ``sensitive'' information from data that is aggregated into training datasets.
This presents the difficulty of determining what information is ``sensitive;'' sensitivity depends on context and there is no generally correct heuristic for this procedure~\cite{brown2022does}.
In contrast, the strict confidentiality of encryption is agnostic to message contents, and provides systemic security benefits by protecting mundane communications alongside sensitive ones (as discussed in Section~\ref{sec:background:e2ee:systemic-e2ee-properties}).

\textbf{Federated learning} is another approach intended to enhance the privacy of training data, by having end-users locally train local models with their personal data. The resulting local model information is then sent to and aggregated into a single model by a central entity. The privacy benefit is that the raw personal data does not leave users' devices. But the local models, which are dependent on that personal data, \emph{do} leave users' devices. In turn, the final centralized model is still dependent on users' personal data as it depends on these local models. Thus, federated learning does not address the issues of training data leakage through model use or adversarial attacks. %

\textbf{Multi-party computation} (MPC) is a technique that allows multiple parties (e.g., users or companies), each with a private data set, to collectively compute a function (e.g., model training) on all of their data sets combined and access the result (e.g., a trained model), without directly sharing their data sets with each other. 
To use MPC, all parties must agree on a function they want to compute. MPC provides the guarantee that parties will learn the desired function output (e.g., the trained model) and \emph{nothing more than that output}.\footnote{Necessarily, by learning the output, parties may also learn information that can be inferred from that output.} MPC does not constrain the function parties choose to compute, and in particular, allows for the function output to be dependent on, and reveal information about, the input data sets---such as when the output is a trained model.\footnote{The confidentiality guarantee offered by MPC thus depends heavily on the function that the parties choose to compute, which they can choose freely. For example, if the function they choose to compute is the identity function---i.e., the output is simply a copy of all of the parties' data sets---then MPC provides no confidentiality at all. If the chosen function consists of aggregate statistics---e.g., the average salary of the many employees in an industry---then MPC provides a stronger guarantee.} A trained model resulting from an MPC computation is just as dependent on its training data as a model trained without MPC. Thus, MPC does not address the issues of training data leakage through model use or adversarial attacks.

\medskip

All of the above techniques provide valuable privacy protections. However, their privacy goals are different in kind from the privacy goals of E2EE systems. As such, they cannot achieve E2EE confidentiality for training shared AI models on E2EE content. Future advances in the technologies discussed will likewise be valuable but will not bring these technologies closer to E2EE, because the type of privacy protection they provide is different in kind from E2EE confidentiality.
Put differently, private training techniques create a \emph{spectrum} of privacy---as determined by, for example, the level of noise introduced, or how many sensitive words are removed from the message---whereas E2EE is a \emph{binary} kind of guarantee.

Private training techniques themselves provide a lens to illustrate the tension between E2EE and AI training: we can frame the guarantees of E2EE as ``extreme'' forms of various private training techniques, which correspond to degraded, impractical instantiations of each technique. For example, the guarantees of E2EE can be cast as differential privacy with $\varepsilon = 0$ (i.e., when the statistical distance between training data with or without a particular item included is 0). However, this value can only be achieved when private data is removed from training to begin with~\cite{wutschitz2023rethinking}, at which point differential privacy is vacuous. Similarly, the guarantees of E2EE can be described as data sanitization in the case when all parts of an input are considered sensitive. Thus, sanitization techniques would remove E2EE messages entirely, which is equivalent to not training at all to begin with. As these examples show, to meet E2EE security, private training techniques essentially degrade to cases where training data is removed altogether.

\subsection{How AI Features are Integrated Matters for E2EE}
\label{subsec:ai_integration}
At time of writing, messaging applications with integrated AI assistants typically offer multiple ways of interfacing with AI features.
These differences may have
implications for E2EE compatibility.
For example, at time of writing, WhatsApp offers at least three distinct ways to invoke Meta AI.
\begin{enumerate}
    \item Through the search bar, which is labeled ``Ask Meta AI or Search.'' When the search bar is selected, it previews some example prompts for Meta AI.\footnote{The search bar can be used either to locally search a user's WhatsApp conversations or to ``Ask Meta AI,'' and appears at the top of a user's conversation list.}
    \item Through a button with the Meta AI logo at the bottom right of the Chats page, which takes you to a one-on-one chat with Meta AI.
    \item Through typing ``@Meta AI'' in a chat with other users. When typing the ``@'' symbol, ``Meta AI'' also appears as a recipient like any user in a group chat, and is put at the top of the list.
\end{enumerate}

It is the third means of invoking Meta AI, which is positioned within the messaging service---and at times, seems to present Meta AI much like another ``user'' or ``contact'' within a user's conversations and group chats---which falls squarely within the scope of our concerns and recommendations.

The first way of invoking Meta AI is a \emph{standalone} AI feature (as defined in Section~\ref{sssec:standalone-vs-integrated}), because only user-initiated search-bar queries that explicitly invoke Meta AI are processed by Meta AI, without involving any E2EE content---and thus is outside the scope of our concerns. %

The second way of invoking Meta AI is also, technically speaking, a standalone feature not integrated with E2EE content. This is because the only content accessible to the AI is in dedicated user-initiated interactions or ``chats'' between the user and Meta AI (and nobody else)---in effect, a chat-like user interface for a direct user-AI interaction that involves no AI processing of E2EE content,\footnote{We consider content in E2EE conversations \emph{between users} to be E2EE content, and distinguish this from direct interactions between a single initiating user and an AI assistant, which we do not consider to contain any E2EE content even if packaged in a chat-like interface. This is consistent with our definition of E2EE content in Section~\ref{sec:defs:e2ee-terms}.}  just like the search bar. However, this chat-like interface may create potential for confusion.
As noted in Section~\ref{sssec:standalone-vs-integrated}, standalone AI features are generally outside our scope, \emph{unless} the provider creates the impression that the AI feature is itself E2EE. Because of ambiguities in the claims that Meta makes about whether ``chats'' with Meta AI are E2EE, the one-on-one chats with Meta AI \emph{are} within the scope of our recommendations.\footnote{
Section~\ref{sec:recommendations:discosure-consent} provides more detailed discussion of the ambiguities.}

\section{Background II: Consent, Privacy, Users, and Areas of Law}
\label{sec:background-legal}
\label{sec:background-consent}

Next, we turn to legal and normative considerations around clear disclosures about and meaningful consent to the use of AI integrated with E2EE systems, complementing our preceding analysis of when and how such systems can be technically compatible with the confidentiality promised by E2EE. 
The legal landscape around E2EE and AI is rapidly evolving, raising questions about user protection and data practices that, in turn, have implications on security and confidentiality. This section provides preliminary background on data practices (Section \ref{sec:socio:what-users-understand}), notice and consent (Section \ref{sec:sociotechnical:notice-consent}), and relevant areas of law (Section \ref{sec:sociotechnical:areas-of-law}).

\subsection{What Users Understand About How Their Data is Used}
\label{sec:sociotechnical:dark}
\label{sec:socio:what-users-understand}

Users are typically unaware of data practices and don't often know what data is being collected about them, by whom, and for what use~\cite{vankleek2017}. Privacy issues are often context-dependent, and decisions may be made based on how safe or trustworthy a company is perceived to be~\cite{pew2016privacyinfosharing}.  

\paragraph{TOS and privacy policies in practice}
When evaluating whether to try a new product, users cannot easily understand terms of service (TOS) and privacy policies, which are often incomprehensible in practice, exceeding reasonable standards of length and readability \cite{nyt-privacy-policies}.  Research has shown that very few people read privacy policies and terms of service, and that reading these for all the apps that a person uses would take hundreds of hours a year \cite{MC09}. Even if users did somehow read all the available information about how their data is being handled, it would be unrealistic to expect them to assimilate this information and act in their best interest based on it, given the vast amounts of data generated over an extended period of time \cite{acquisti2005}. 

\paragraph{Deceptive patterns}
Another barrier to users' ability to make informed decisions about their data is the use of deceptive design. These deceptive patterns (or ``dark patterns'') are sets of design choices that coerce users to do things that aren't in their best interest. Dark patterns can take the form of trick wording, comparison prevention, hidden costs, and nagging. These techniques can cause financial harm, loss of control (for instance over data usage), and discrimination \cite{brignull-2023}. %
In the case of privacy policies, dark patterns may be employed to coerce users into making choices that are beneficial to the given service, but not necessarily beneficial to the user. 
Deceptive patterns are often used to make users part with more data than they may be aware of or be willing to disclose \cite{nouwens2020}. 
Documenting and regulating dark patterns is challenging: they are deceptive, so users often are not aware of them, and users may also feel embarrassed about having been tricked, which complicates reporting \cite{iobscura}.

\paragraph{Data can be surprisingly revealing}
It is possible to infer surprisingly invasive information from someone's shopping habits; even to predict a consumer's pregnancy before close family members are made aware \cite{NYTprivsecrets} \cite{Forbestargetpreg}. Time-stamped record of an individual’s cell phone location may be stored by mobile network operators for extended periods of time without the user’s knowledge.\footnote{\textit{Carpenter v. United States}, 585 U.S. 296 (2018)}. 

\paragraph{Data brokers} Companies that collect data from users often sell user information to data brokers who combine the data and build incredibly detailed dossiers about individuals' profiles that are available for sale to the general public \cite{wireddatabrokers}. Users may not know that data brokers even exist. %

\paragraph{The ``Privacy Paradox''} Consumer actions tend to undervalue their privacy. This social phenomenon is captured by the ``privacy paradox,'' which identifies that people’s stated concerns about privacy are often not reflected in their online behaviors \cite{privbehavecon}. Thus, giving consumers a choice to opt in or out will often not effectively translate their preferences. Additionally, the sheer volume and time constraints of decision-making with regards to privacy has been known to cause ``consent fatigue'' \cite{murkyconsent}. Traditionally, consent has been provided in a binary format, like an on/off switch. However, as services offer wider ranges of settings and process increasingly more types of user data, privacy choices have become more granular. These consent requests compete for consumers limited time and attention, making it less likely for each request to be considered appropriately \cite{WPprivpol}. 

\subsection{The Failure of Consent}
\label{sec:sociotechnical:notice-consent}

\subsubsection{Legal Consent}

The concept of user \emph{consent} plays a significant role in nearly all laws that regulate privacy \cite{murkyconsent}. From contract law to tort law to privacy statutes, notions of consent are relied upon to legitimize a wide range of data collection and processing activities that would otherwise not be permissible. Legal definitions describe consent idealistically as an individual's voluntary agreement that is wilfully and freely manifested, in a manner that is not coerced, fraudulent, or erroneous \cite{wex_consent}.
In practice, the situations in which courts consider a user to have consented to a condition often fail to meet this idealistic description. Normatively, the legal framework of consent has been heavily criticized for insufficiently reflecting the realities of consumers and for ignoring the underlying social dynamics of data privacy.

Legal consent presupposes a prior \emph{disclosure} or legal \emph{notice}. In digital services, legal notice may come in the form of a privacy statement, terms of service, or banner displayed by websites and apps. 
In contract law, privacy notices are regarded by a great majority of courts as a valid contracts, despite extensive research (see Section~\ref{sec:sociotechnical:dark}) showing that these notices are not read in practice \cite{WurglerBarGill}. Terms of service and privacy policies are generally upheld as legally binding contracts even if users did not consent by an affirmative action like signing a document or checking a box labeled ``I agree'' while creating an account---American courts often treat a user's continued use of a website or app as ``implicit consent''  to its terms and conditions.

The ``notice and consent'' (or ``notice-and-choice'') approach is the prevailing framing of consent in the United States. The underlying rationale is that consumers have the choice to stop using the service entirely or to opt out of specific features if they are not in agreement with the provided terms or notice \cite{shapeofconsumercontracts}. 
Laws like the Gramm-Leach-Bliley Act, the Children's Online Privacy Protection Act, and the Health Insurance Portability and Accountability Act, which respectively regulate financial information, children's data, and health data, require organizations to provide individuals with a privacy notice and an opt-out mechanism. 

Another legal approach to consent is to require the ``express consent'' of users. This approach requires consumers to opt in by taking an affirmative action to indicate consent, such as by checking a box or clicking an ``Accept'' button. The European Union’s General Data Protection Regulation requires ``informed,'' ``specific,'' and ``unambiguous'' consent, where the term ``unambiguous'' is understood to reject notions of implied consent through inaction.\footnote{Regulation (EU) 2016/679 ("GDPR"), Article 7 and Recital 32.} Due to the European Union's regulatory strength, the so-called Brussels effect, and the GDPR's exterritorial scope, many privacy laws across the globe have adopted similar standards for consent \cite{brusselseffect}. In the United States, although the notice-and-choice approach is still more prevalent, some U.S. state privacy laws require express consent in specific circumstances.\footnote{E.g., the Colorado Privacy Act requires opt-in consent for processing sensitive data. (CPA § 6-1-1308 (7))} 

\subsubsection{Normative Critiques}
\label{sec:normativecrit}

Regardless of the consent regime adopted by privacy laws, the conditions for meaningful consent are widely regarded as fictional in normative terms. Meaningful consent would be informed, not coerced or unduly manipulated, and require that individuals have the capability to make an appropriate risk assessment (e.g., based on a cost-benefit analysis) \cite{murkyconsent}. 

There are many reasons why users are unable to meaningfully consent, many of which are related to the considerations already discussed in Section~\ref{sec:socio:what-users-understand}. Notoriously, notice and consent is inadequate in informing individuals about an organization's privacy practices because consumers do not read privacy notices \cite{Wurgler_Disclosure}. %

Moreover, understanding data privacy implications often requires technical expertise. This is especially true for AI/ML systems, which are frequently opaque and referred to as ``black box'' models. Even the developers of many such systems do not fully understand the privacy implications of training their models on user data \cite{nobodyknows}. In general, consumers cannot be expected to understand the implications of disclosing particular pieces of their data, especially in aggregate forms.  Consumers overestimate baseline legal protections \cite{privhomoecon} and underestimate the risk seemingly inconsequential disclosures pose over time \cite{SoloveDilemma}.

Regarding the contractual relationship between consumers and online platforms, consumers do not have meaningful bargaining power or choice \cite{cohenexamined}. The monopolistic tendencies of online social networks are exacerbated by the low costs of scalability and increased value as the number of users grow, resulting in lack of real choice between services \cite{cohenplatecon}. Even if alternative services exist, privacy practices and terms will often be the same among competitors. Additionally, the asymmetric nature of the relationship means that individual opt-outs require scale in order to exercise compelling power against large corporations.
    
Another often overlooked condition of consent is its validity across time. Viewing consent as a one-time deal does not match the ongoing nature of data processing activities. Consent may expire or may simply not be enough to capture an activity that happens too far into the future, where circumstances may have changed for either party. Attempting to solve this, some laws like the GDPR and the Colorado Privacy Act explicitly acknowledge that consent may be withdrawn at any time and that consumers have a right to request for data deletion.\footnote{Regulation (EU) 2016/679 ("GDPR"), Article 7(3) and Article 17(1)(b) and CPA ("Colorado Privacy Act") §6-1-1306 (1)(a)(IV)(C).} Although these are technically two different user rights, in the case of recurring data processing activities, the withdrawal of consent will often require deleting personal data in order to honor the consumer's wishes. However, considering the ongoing nature of AI/ML system-training, the deletion of data is, in general, infeasible.  Thus, opting out or withdrawing consent after the data has already been used for training is rendered pointless and ineffective to stop undesired results, such as a model outputting the personal data on which it was trained.

Finally, legal consent largely elides the social dimensions of privacy. Although privacy can easily be understood to have a social connotation that highlights its value as a shared good, legal consent (in its current form) cannot, and is typically linked to an individual permission. A person is unable to consent for another unless in some legally accepted representation capacity.  Multi-party messaging data raises complex issues outside the scope of existing legal consent frameworks, as legal consent does not account for the fact that shared data is extra-individual by nature. Genetic data is another example of information in this category, as it reveals information about a person's entire family.

\subsubsection{Social Dimensions of Information Privacy}
\label{sec:sociotechnical:social-dimensions}

In his influential book \textit{Code and Other Laws of Cyberspace}, Larry Lessig argued that laws are only one kind of the several regulations that affect the way people behave~\cite{codeandcyberspace}. Architecture (whether of buildings or code), markets, and social norms can regulate individuals' behavior just as much, if not more than, law. 

Defining social norms precisely may be an impossible task in its own right. Helen Nissenbaum's theory of contextual integrity argues that the term ``privacy'' must be understood depending on the context~\cite{contextualintegrity}. The norms around privacy in a health context may be completely different than in a school context, and from group to group. For example, if a child is diagnosed with ADHD by a private psychiatrist, it would be a severe violation of the child's privacy for the psychiatrist to disclose that information outside of the child's family. However, if the psychiatrist is employed by the child's school, disclosure to the child's teacher and the school's accommodation staff may be expected. Blanket consent fails to account for these norms and undercuts privacy—addressing the flow of data itself is the only way to fully address privacy concerns.

Linnet Taylor, Luciano Floridi, and Julie Cohen are three of many academics who argue for notions of ``group privacy." The motivating thought behind this concept is that giving each individual the last word on privacy concerning their personal data may not capture all the interests at stake. There are different flavors: Taylor and Floridi argue that informed consent can only be given by a data subject, so if the subject is a group, the consent of individuals is meaningless~\cite{groupprivacy}. The definition of the relevant group greatly affects the manner in which informed consent can be given. For example, acquiring the consent of a group chat would look very different than acquiring the consent of an ethnic group. Cohen argues that reduced privacy harms the normal functioning of society in a way that manifests far beyond the individual~\cite{whatprivacyisfor}. For example, individualized search engine results distort citizens' perception of the world and make participation in the democratic process less likely to result in compromise. Both of these approaches acknowledge that an individual's grant or withholding of consent is unlikely to capture the full set of issues involved.

Several scholars have proposed changes to consent based on the idea of group privacy. August Bourgeus and Laurens Vandercruysse suggest a form of semi-automated consent, which would allow individuals to set their preferences accurately once and carry them throughout their online life~\cite{reinvigoratingconsent}. Daniel J. Solove advocates for ``murky consent,'' a baseline model which would rely on data holders accepting a duty of loyalty, duty to avoid thwarting reasonable expectations, and a duty to avoid unreasonable risk~\cite{murkyconsent}. Jack Balkin, and many scholars since, argues that holders of sensitive information should act as fiduciaries to data subjects~\cite{informationfiduciaries}.

\subsection{Areas of Law}
\label{sec:sociotechnical:areas-of-law}
We refer to several substantive areas of law that affect E2EE service providers. They are summarized below.

\paragraph{Contract law}
Contracts are agreements between private entities. ``Contract law'' is a legal framework that enables enforcement of contracts by the judicial system. With few exceptions (for illegal and unconscionable conduct), contract law does not place limits on what may be included in a contract. This ``freedom to contract'' is especially strong in the United States.

\paragraph{Consumer protection law}
Consumer protection law places limits on certain practices that corporations might undertake that could harm consumers. 
Thus, consumer protection law can be seen as one limitation on the freedom to contract. 
For example, one particularly voluminous subfield of consumer protection law encompasses all the safety regulations regarding car design.

\paragraph{Data protection law}
Data protection laws are regulations that govern how personal data is handled and used. The definition of personal data may vary between jurisdictions, but generally means information that can identify individuals.  These laws establish rules and guidelines for lawful processing of personal data. 

\paragraph{Antitrust law}
Antitrust laws are regulations that promote competition and prevent monopolies by regulating the conduct and organization of businesses. 

\section{Legal Implications of Integrating AI with E2EE}
\label{sec:sociotechnical-implications}
\label{sec:legal-implications}

The integration of AI assistants into applications that process any form of E2EE data raises important considerations under current legal regimes, which are related to the security considerations we have laid out thus far, as well as broader legal issues around disclosure and consent.

In this section, we begin with a brief survey of privacy terms in several E2EE messaging service providers' Terms of Service and Privacy Policies (Section~\ref{ssec:tos}).
We then overview relevant legal obligations under data protection and consumer protection law, considering how AI service providers are currently regulated in the US (Section~\ref{ssec:us-legal-considerations}) and in the EU (Section~\ref{ssec:eu-legal-considerations}). 
The specific issue of AI processing of E2EE data has not yet come up in legal proceedings, but our discussion provides background on relevant existing law, as well as ongoing controversies and uncertainties, that will inform how AI processing of E2EE data is treated under US and EU law.
Finally, we note regulatory trends across other jurisdictions (Section~\ref{ssec:other-jdx}) and provide a brief note on how competition law may affect data-driven business operations (Section~\ref{ssec:antitrust}).

\subsection{Terms of Service and Privacy Policies of E2EE Services in Practice}
\label{ssec:tos}

When drafting terms of service and privacy policies, international companies must account for several significant regulations. 
In general, companies are incentivized to write overly broad terms of service that cover any possible future business development, because courts generally uphold broad contract terms, especially in the United States~\cite{Wurgler_Disclosure}. In the case of an E2EE messaging service, for example, a company might be incentivized to include a term in its fine print that says that it may stop offering E2EE at any time, even if all of its promotional material touts its E2EE, and even if it does currently offer strong E2EE and has no plans to cease providing it.
Similarly, while E2EE messaging services may have high standards with regards to the third parties they contract with, they regularly attempt to disclaim liability for the actions of any other entity. Third party liability disclaimers are found in all four of the terms of service summarized below.\footnote{This does not prevent companies from including quality assurances in contracts with vendors. In the event of a data breach, vendors can be liable to both direct customers and end consumers~\cite{inreblackbaud}.} So, while E2EE messaging services may advertise lofty ideals, and even if they really do promote those ideals by deploying strong privacy technologies, sometimes their terms of service may not make the same promises. For example, WhatsApp's terms of service would allow it to share user data for marketing purposes, a function that would conflict with the E2EE that it \emph{does}, elsewhere, commit itself to providing. Such inconsistencies are common in terms of service because of the incentive to over-include.

The contractual privacy terms of four messaging services which offer E2EE features are summarized in Table~\ref{tab:e2eeprivacypolicies}. Notably, the three services (WhatsApp, Meta's Messenger, and Telegram) that do not promise use of E2EE do not use it for messages with bots or businesses. Bots or businesses could store user messages in plaintext if they chose to---in other words, these three services have decided not to dictate third-party message handling practices, instead allowing the third-party's privacy policy to govern these interactions. Finally, WhatsApp, Telegram, and Signal broadly market themselves as ``secure'' messaging platforms, and occasionally use the term ``end-to-end encryption'' in marketing. Meta's Messenger, on the other hand, does not mention security on its splash page, at the time of writing.

Of the four services discussed, WhatsApp's privacy policy notably promises less security than its marketing materials promise, and than its product actually provides. To our knowledge, the service as provided at the time of writing falls into Section \ref{sec:buckets}'s Category 2, consistent with marketing materials, while the privacy policy describes a Category 3 system. 

\begin{table}[h]
    \small\centering
    \begin{tabularx}{\textwidth}{>{\centering\arraybackslash}X>{\centering\arraybackslash}X>{\centering\arraybackslash}X>{\centering\arraybackslash}X>{\centering\arraybackslash}X} 
    \toprule
         \textbf{Service} & \textbf{Legible privacy policy in a single document?} & \textbf{E2EE Promised?} & \textbf{Circumstances of transfer of consumer data to third parties}& \textbf{Limitations to third parties on-platform mentioned in ToS}\\ 
         \midrule
         \textbf{WhatsApp}&  Yes&  ``Offered''&  For marketing, normal operations& None\\ 
         \textbf{Meta's Messenger}&  No - spread across complex privacy center&  ``On some Products, you can use end-to-end encrypted messages.''&  For marketing and to advertisers on Facebook, normal operations& None\\ 
         \textbf{Signal}&  Yes&  Yes&  For text message verification& N/A\\ 
         \textbf{Telegram}&  Yes&  For ``secret chats''&  For on-platform purchases, translation, and transcription& Bots can only get username, language, IP, and shared messages\\ 
         \bottomrule
    \end{tabularx}
    \caption{Comparing Privacy Policies of Select E2EE Messaging Services}
    \label{tab:e2eeprivacypolicies}
\end{table}

\subsection{Overview of US Legal Considerations for E2EE \& AI}
\label{ssec:us-legal-considerations}

Data protection statutes provide consumers with relatively limited protections as well. The United States does not have a federal general data protection law and sector-specific laws would only cover messaging services in the rarest edge cases. While some states, like California, have stronger general data protection laws, services can and do vary their products from state to state to comply with these laws. The GDPR may have more significant implications, although its scope is also limited, and many general data protection laws outside of the EU are modeled after the GDPR and suffer from the same limitations. 

In the US, consumer protection law, a broad hammer mostly wielded by the FTC and state attorneys general, is likely to be provide stronger protections than data protection law, and has the flexibility to provide strong checks against the most egregious business practices, as discussed next.

\subsubsection{Consumer Protection and the US Federal Trade Commission}

Consumer protection law in the United States would constrain some implementations of AI in E2EE systems. At the federal level, the Federal Trade Commission Act of 1914 (15 U.S.C. § 45) created the Federal Trade Commission (FTC) and enabled it to investigate ``unfair'' or ``deceptive'' trade practices. 

Deceptive trade practices---the most relevant category here---are defined as those which constitute ``a misrepresentation, omission, or other practice, that misleads the consumer acting reasonably in the circumstances, to the consumer's detriment''~\cite{ftcdeceptionletter}. 
The FTC investigates security-related incidents as deceptive practices. Such investigations are often prompted by data breaches and result in the FTC discovering misleading statements made to consumers about security. The FTC has discretion over which investigations to pursue---and must make choices, as it does not have the capacity to pursue every lead it encounters.

For example, increased public use and scrutiny of Zoom in 2020 led to the discovery that Zoom was not protecting all video calls with the ``end-to-end, 256-bit encryption'' it advertised. The company settled with the FTC, agreeing constructively to implement a comprehensive security program and cease misrepresentations related to encryption~\cite{consentdecreezoom}. 
A dental software company advertising its use of ``encryption,'' unqualified, was found in an FTC proceeding to be misleading customers because the level of encryption used did not meet industry standards~\cite{dentalsettlement}. 
A mortgage company settled with the FTC for advertising that the data it collected was encrypted, when in reality it was encrypted in transit but not after receipt~\cite{mortgagesettlement}.

There is no precise technical specification that companies must meet to avoid being prosecuted for deceptive practices.\footnote{But there is a roughly defined floor, below which companies can be investigated for \textit{unfair} practices. This comes up often when companies are subjects of a data breach compromising unencrypted credit card, passport, or social security numbers~\cite{mariottstarwoodbreach}.} 
That said, under FTC precedent, there are clear lines that an E2EE messaging app could not cross without being found deceptive. If an app is marketed as using ``end-to-end encryption,'' established FTC precedents demonstrate that the app must use end-to-end encryption throughout the processing of the relevant data~\cite{consentdecreezoom,mortgagesettlement},
have E2EE on by default~\cite{consentdecreezoom}, and meet industry standards~\cite{dentalsettlement}. In other words, the marketing of a service in Category 5 (No E2EE) defined in Section \ref{sec:buckets} with an unqualified claim that the product uses ``end-to-end encryption'' would be considered deceptive under FTC precedent.
We stress, again, that the FTC has discretion over which cases it pursues and prioritizes---so despite the above, an app in violation of the above would not necessarily be investigated or suffer consequences.

The FTC is not the only governmental agency capable of investigating deceptive practices. Many state attorneys general have similar powers granted by local versions of the FTC Act.\footnote{For example, the New York State Attorney General investigates deceptive practices under the Deceptive Practices Act, N.Y. Gen. Bus. Law § 349.} State attorneys general are important consumer protection enforcers because they can address smaller or more local concerns that the FTC would miss or lack the bandwidth to address. They can also be significant regulators in their own right, and collaborate between states to take on the most sophisticated corporations. With uncertainty around the enforcement objectives of the FTC during the second Trump administration~\cite{ferguson-ftc}, state attorneys general are likely to take on even more importance as enforcers of consumer protection laws.

\subsection{Overview of EU Legal Considerations for E2EE \& AI}
\label{ssec:eu-legal-considerations}

\subsubsection{EU Legal Background}

In many ways, legal obligations for data-driven services in the EU extend beyond the disclosure and consent model described above. Next, we overview key relevant EU legislation.

\paragraph{GDPR} The GDPR sets out key principles that impose additional constraints upon collecting, storing, and using personal data.\footnote{For example, the ``lawfulness, fairness and transparency'', ``purpose limitation'' and ``data minimization'' principles are important guardrails enshrined in the GDPR. Regulation (EU) 2016/679 (``GDPR''), Article 5(1).} Under the GDPR, obtaining consent is not the only basis to legitimize the processing of user data---it is just one among six. Many user data-related activities within companies can rely on the broad alternative basis of ``legitimate interests,'' that is, where processing is necessary for a ``legitimate interest'' pursued by the organization.\footnote{Regulation (EU) 2016/679 (``GDPR''), Article 6(1)(f).} For instance, analyzing customer data for fraud prevention and signs of criminal activity or engaging in direct marketing with preexisting clients are typically deemed activities within a company's legitimate interests~\cite{icolegint}. The legal basis of ``legitimate interests'' is often seen by organizations as more flexible and less cumbersome than obtaining user consent.\footnote{Despite the requirement to ensure that organizational interests do not override users' fundamental rights (known as the ``balancing test''). Regulation (EU) 2016/679 (``GDPR''), Article 6(1)(f).} 

This is because the GDPR sets higher standards for user consent: it must be ``freely given,'' ``specific,'' ``informed,'' ``unambiguous,'' and may be withdrawn (or opted out) at any time. Employing dark patterns to obtain user consent or to hide mechanisms to opt out may violate the GDPR and other overlapping regulations, such as the 2022 Digital Services Act (DSA) and the Unfair Commercial Practices Directive (``UCP'' Directive) \cite{ECconsumanip}. The European Data Protection Board (EDPB) provided examples of deceptive techniques used by social media platforms that do not qualify as ``freely-given'' consent \cite{edpbdarkpat}. These include ``forced'' consent, where refusing to consent leads to a denial of the service, and ``bundled'' consent, when consent is bundled with the acceptance of the platform's terms and conditions.  These heightened protections are valuable; at the same time, the GDPR still offers an imperfect regime for protecting user rights and falls short of the ideals of freely given, unambiguous, informed consent based on mutual assent (e.g., \cite{cofoneconsent}).

The GDPR is just one piece of data protection law in the EU. Other data protection legislation does mandate obtaining user consent in certain situations, as discussed next.

\paragraph{ePrivacy Directive}
The ePrivacy Directive (Directive (EU) 2002/58 on Privacy and Electronic Communications) requires obtaining user consent for the surveillance of private communications\footnote{Directive (EU) 2002/58 on Privacy and Electronic Communications (``ePrivacy Directive''), Article 5(1).} and the storing of non-essential browser cookies and other tracking technologies in user devices---a requirement that has reshaped the appearance of websites with the display of consent-collecting ``cookie banners''.\footnote{Directive (EU) 2002/58 on Privacy and Electronic Communications (``ePrivacy Directive''), Article 5(3).} Article 5 of the ePrivacy Directive establishes a general prohibition on storing information or gaining access to stored information on users’ end devices without obtaining prior consent, including listening, tapping, or other kinds of interception or surveillance of communications and related metadata. 

As the ePrivacy Directive was originally designed to regulate traditional telecom operators, it is not completely clear whether it applies to ``over-the-top'' (OTT) communication services such as WhatsApp, Telegram and Zoom. 
In 2017, the European Commission adopted a proposal to replace the Directive with 
an updated ePrivacy Regulation 
that would explicitly cover OTT services \cite{EPeprivacy}. Attempts to advance the reform received significant pushback from industry players, and were undercut by the passing of a temporary derogation that allows the scanning of private communications to detect online Child Sexual Abuse Material \cite{edriepriv}.  After being blocked for several years due to complicated trilogue negotiations in the EU Parliament, the ePrivacy Regulation proposal was withdrawn by the EU Commission in February 2025 \cite{EPeprivacy} \cite{EU-privreg-ditched}.

Despite the uncertainty facing potential reforms, the European Data Protection Board (EDPB) recommends that a new ePrivacy Regulation must under no circumstances lower the level of protection offered by the current ePrivacy Directive, as the confidentiality of electronic communications is a fundamental right that requires specific protection  \cite{edpbepriv}. The EDPB further recommends that any exceptions must be narrowly tailored, and should not lead to the systematic monitoring of electronic communication content, nor allow providers or third parties to circumvent any encryption. 

\paragraph{AI Act}
The EU’s recently enacted AI Act reinforces the applicability of both the ePrivacy Directive and the GDPR to AI systems, highlighting that these laws establish the basis for sustainable and responsible data processing.\footnote{Regulation (EU) 2024/1689 (``EU AI Act''), Recital 10 and Article 2(7).} Recital 69 of the AI Act notes that the rights to privacy and to protection of personal data---which are fundamental rights in the EU\footnote{Charter of Fundamental Rights of the European Union (2000), Articles 7 and 8.} ---must be ``guaranteed throughout the entire lifecycle of [an] AI system,'' and that measures taken by providers to ensure compliance may include anonymization, encryption, and ``the use of technology that permits algorithms to be brought to the data and allows training of AI systems without the transmission between parties or copying of the raw or structured data themselves''.

\subsubsection{Training AI in the EU and the UK: Ongoing Controversies}

Conditions for processing user data to train AI have been a contentious topic in the European Union and the United Kingdom---even for user data in technically public posts, let alone in more private contexts. Throughout 2024, tech companies like Google, Meta, X and LinkedIn have delayed or suspended plans to launch generative AI models in the EU due to regulatory inquiries into their training practices \cite{politico}.  So far, these companies have been aiming to harness user-generated content to train their AI assistants or large language models and have not been openly targeting private messaging. Nevertheless, AI training practices have already been raising substantial privacy concerns in the EU.

Ireland's Data Protection Commissioner (DPC) has actively monitored the release of several AI-powered virtual assistants fueled by EU user data. The Irish authority is the European Union's lead regulator for many US tech companies due to their EU headquarters being based in Dublin. In June 2024, the DPC raised concerns about Meta’s privacy policy update that allowed Meta AI to be trained on public posts of EU users on Facebook and Instagram \cite{dpcmeta}. The DPC acted in response to a complaint filed by the advocacy group NOYB - European Center for Digital Rights.
NOYB filed complaints in eleven countries urging regulators to act against Meta’s ``abuse of personal data for AI'', drawing attention to the feature’s lack of opt-in consent and hidden opt-out form \cite{noybmetaai}.

Similarly, in August 2024, the DPC initiated proceedings against X (formerly Twitter) in the Irish High Court for concerns related to the training of X’s AI assistant Grok \cite{dpcgrok}. A few months prior, users noticed a default setting on the app, collecting user consent via a pre-ticked box, allowing X to use posts, interactions, inputs and results for AI training \cite{guardiangrok}. The proceeding was dismissed after X agreed to suspend the use of publicly posted data on X to train Grok. In September, the DPC opened an inquiry into Google’s use of European Union users' personal data to help develop its foundational AI Model, Pathways Language Model 2 (PaLM 2) \cite{reuterspalm}. The DPC also publicly confirmed that it has been in contact with Apple regarding its plans for Apple Intelligence AI \cite{ieapple}.  At the broader industry level, an open letter coordinated by Meta and signed by executives from more than two dozen companies has criticized the EU’s regulatory approach for generating uncertainty and for blocking innovation in the region \cite{wsjmeta} \cite{metaopenletter}. 

Other regulators have taken slightly different approaches. In contrast to the EU, plans to launch Meta AI were resumed in the United Kingdom after Meta’s engagement with the UK Information Commissioner Office’s (ICO) \cite{metauk}. Under the UK GDPR,\footnote{Following Brexit, the UK retained a copy of the GDPR in domestic law, which is known as the UK GDPR.} Meta confirmed it will be relying on ``legitimate interests''---not user consent---as a legal basis to train Meta AI on the public posts and comments of users above the age of 18.  The ICO mandated certain measures from Meta, including making it simpler for users to object to the data processing and providing them with a longer window to do so, as well as providing more transparency about how the data is being used, and highlighted that it did not provide final regulatory approval and that Meta is bound to demonstrate ``ongoing compliance'' \cite{icometa}. Meta confirmed it will be sending in-app notifications to inform users about the feature and maintains that private messages are not within the scope of training data \cite{metaaieurope}.

Regulator inquiries and guidelines from EU member states have typically operated on a case-by-case basis, taking into consideration each AI system’s business model and operational specifications. The European Data Protection Board (EDPB), which oversees the consistent application of data protection rules across the EU, began harmonizing the GDPR's interpretation regarding AI training after the launch of ChatGPT. Noting that the Italian regulator took enforcement action against OpenAI, Inc. due to several GDPR violations, the EDPB launched a dedicated task force to investigate privacy issues in the ChatGPT service \cite{edpbchatgpt}. The main findings concerning AI training were based on OpenAI’s practice of web scraping (the automatic extraction of  data from third-party platforms). This practice is different from obtaining data posted by users on AI companies' own platforms, as Meta AI and Grok’s processing operations currently do. Nevertheless, some of the EDPB’s recommendations may be more broadly applicable beyond the specific case of OpenAI and ChatGPT, such as the requirement that users be clearly and demonstrably informed that any prompts, including text queries and file uploads, may be used for training purposes.

Aiming to trigger more comprehensive discussions at the supranational level, the DPC requested the EDPB for a supervisory opinion on relevant issues when processing personal data to develop and train AI models \cite{dpcgrok}.  The EDPB's much-awaited Opinion 28/2024 was published in December 2024,   focusing specifically on (1) when and how an AI model can be considered as ``anonymous"; (2) how organizations can demonstrate the appropriateness of legitimate interest as a legal basis in the development and (3) deployment phases; and (4) consequences of the unlawful processing of personal data in the development phase of an AI model on subsequent processing operations \cite{edpbdecai}. With regard to balancing organizations' ``legitimate interests'' and user rights, the EDPB highlighted the importance of transparency and meeting users’ reasonable expectations considering the complexity of the technologies and the variety of potential uses of personal data. The EDPB did not address user consent; the EDPB noted that the opinion does not cover all EU data protection issues in the context of AI development and deployment.

Importantly, while the EU is openly debating the lawfulness of AI training under different conditions, none of the regulator inquiries or opinions to date address the issue of using data obtained from messaging or end-to-end encrypted services. While data posted on social media platforms under public settings may generally be considered public, messaging data is private by nature.\footnote{There are also nuances regarding the public nature of publicly posted data: ``(...) the mere fact that personal data is publicly accessible does not imply that 'the data subject has manifestly made such data public'. In order to rely on the exception laid down in Article 9(2)(e) GDPR, it is important to ascertain whether the data subject had intended, explicitly and by a clear affirmative action, to make the personal data in question accessible to the general public.'' See \cite{edpbchatgpt}.} Legal regimes typically recognize that private communications must be treated to a higher standard of privacy protection because of their more intimate and vulnerable qualities, as embodied by the EU’s ePrivacy directive.\footnote{In the US, the Electronic Communications Privacy Act of 1986 protects wire, oral, and electronic communications from unauthorized interception, disclosure, or use, including email, phone calls, and data stored electronically.} Thus, the use of private messages shifts the paradigm and necessarily calls for more comprehensive debates around user consent. Moreover, training on what is advertised or commonly thought to be E2EE data may frustrate users' legitimate expectations as to the privacy and security of their data.

Although the regulation of AI training in the EU is still in flux and conclusions are uncertain, the recent inquiries in the EU and the UK have indicated that simply updating the company’s privacy policy to allow AI training is insufficient to comply with GDPR standards. Besides providing users and regulators with proactive transparency and disclosure, the substance of how companies are collecting personal data, the types of data harnessed and how they are being used also matter. If private messages are used as training data, questions related to the indispensability of user consent, adequate interfaces for collecting consent and options to exercise control over data become substantially more pressing. 

\subsection{Other Jurisdictions}
\label{ssec:other-jdx}

The EU GDPR is seen by some as the gold standard for online privacy protection \cite{EDPS}. 
Multiple jurisdictions have enacted new regulations or updated existing privacy laws, many of which emulate the GDPR’s provisions, indicating a global trend to increase data regulation. Due to the GDPR's influential role, European regulators' stance on AI training issues will likely trigger ripple effects across the globe \cite{brusselseffect}. 

For instance, following the Irish regulator's reaction to Meta AI, Brazil’s data privacy regulator (National Agency for Data Protection (ANPD)) similarly suspended Meta’s privacy policy update allowing the training of Meta AI on the data of Brazilian users \cite{anpdmeta}. Brazil’s first comprehensive data protection regulation (the ``LGPD'') was enacted in 2020 and is heavily inspired by the GDPR. Brazil is reported to be WhatsApp's second biggest market worldwide, with more than 124 million active users \cite{emarketer}. The ANPD raised concerns as to Meta's legal basis for processing sensitive and non-sensitive data, transparency in how the changes in the privacy policy were communicated, and accessibility to exercise the right to opt out. 

The regulator's technical report also noted discrepancies in Meta's standards for EU and Brazilian users, despite similarities between the GDPR and the LGPD. Unlike EU users, Brazilians were required to go through eight steps to access an opt-out form hosted in a complex interface \cite{fpfbrmeta}. Additionally, while EU users were notified about the privacy policy update via email and app notifications, Brazilian users were not directly informed and were only able to access the update via Meta’s Privacy Policy Center. After several rounds of negotiations and subsequent fixes carried out by Meta, the ANPD lifted the suspension \cite{anpdmeta2}. Meta AI was launched in Brazil in October 2024, although the feature is still undergoing full regulatory inspection.

Meta has recently faced severe enforcement actions for privacy violations in other jurisdictions that have updated their privacy laws to be more robust. In November 2024, South Korea’s privacy regulator fined Meta 21.6 billion won (roughly 14 million euros) for illegally collecting sensitive personal information from Facebook users, including data about their political views and sexual orientation, and sharing it with thousands of advertisers without users' express consent \cite{euronewssk}. In July 2023, Meta subsidiaries were fined by the Australian Competition and Consumer Commission a total of \$20 million Australian dollars for misleading consumers about the privacy protections in Meta-owned VPN app Onavo Protect \cite{accmeta}. Although the app was marketed as a privacy tool in app stores, it gathered extensive user data for Meta's commercial purposes.
 
These enforcement actions show a trend of increased regulatory oversight over digital corporations’ data practices stemming from legal privacy obligations. As corporations are faced with multiple cumulative regulatory standards, they may be encouraged to conform to the combination of the most stringent regulations provided by different jurisdictions \cite{brusselseffect}. However, different jurisdictions will exercise varying degrees of leverage over tech companies based on the company's market share and number of users by country or region, as well as the local law's penalty provisions and the value of the currency.

\subsection{Privacy, Antitrust and Consumer Data}
\label{ssec:antitrust}

Big tech corporations are increasingly at the center of discussions about both antitrust and data protection regulation. Access to vast amounts of consumer data is a strategic asset that may reinforce online platforms’ dominant positions by creating barriers to entry for competitors, reducing consumer choice, or unfairly manipulating markets \cite{cohenplatecon}. Mergers and acquisitions between tech corporations may involve the merging of databases and data sharing practices in violation of privacy laws \cite{icoma}, at times blurring the lines between privacy and antitrust concerns. We briefly provide two illustrative examples.

At the time of Meta’s acquisition of WhatsApp in 2014, the European Commission approved the acquisition \cite{ecmerger1} with the understanding that Facebook would be unable to do automated matching between Facebook users' accounts and WhatsApp users' accounts \cite{ecmerger2}. However, in August 2016 WhatsApp announced updates to its terms of service and privacy policy including the possibility of linking user data across WhatsApp and Facebook. The aftermath resulted in a \texteuro110 million fine on Meta (formerly Facebook) for providing misleading information under EU merger regulation \cite{ecmerger2}. Had a separation between WhatsApp and Facebook data been enforced beyond levying a fine, this antitrust measure could have had far-reaching implications for the integration of certain operations across WhatsApp and Facebook.

The EU's Digital Markets Act (DMA), applicable since May 2023, is a landmark antitrust regulation aimed at curbing Big Tech’s market dominance and ensuring fair competition between digital services. The DMA recognizes privacy and data protection as key principles and specifically regulates the prevention of cross-service data merging in Article 5(2).\footnote{Regulation (EU) 2022/1925  ("Digital Markets Act - DMA"), Article 5(2).} This provision states that gatekeepers cannot combine personal data from different services unless they obtain explicit opt-in consent from users. 

Outside the EU, in November 2024, India’s antitrust regulator targeted Meta for anti-competitive behavior related to data sharing \cite{indiaexpress}. The Competition Commission of India (CCI) imposed a penalty of roughly 25.4 million dollars on Meta after WhatsApp users were forced to to share their data with other Meta-owned companies by means of a 2021 privacy policy update with no opt-out mechanism \cite{indiareuters}. 
The regulator also ordered WhatsApp not to share user data for advertising purposes with other Meta-owned applications for five years, highlighting that the data sharing resulted in denial of market access for competitors.

These examples illustrate how data privacy decisions can lead to broader antitrust concerns beyond the scope of privacy and vice versa. While companies often have monetary incentives to access larger pools of data and integrate features throughout a broad range of products, this may come at the expense of higher scrutiny by regulators. This is a particularly timely observation as the past four years have seen a wave of antitrust lawsuits against Big Tech initiated by the US and the EU \cite{harvardantitrust}.  That said, the 2025 government administration changes in the US may change priorities at the FTC, which is a prominent enforcer of both antitrust and privacy laws.

\section{Recommendations}
\label{sec:recommendations}
This section presents our recommendations for integrating AI assistants into E2EE applications. 

Our recommendations address both compatibility with E2EE from a \emph{security perspective} (Recommendations 1 and 2, discussed in Section~\ref{sec:recommendations:cake}), and disclosure and consent from a \emph{legal and normative perspective} (Recommendations 3 and 4, discussed in Section~\ref{sec:recommendations:discosure-consent}).
We conclude the section with an example application of our recommendations to the practical example of Apple Intelligence (Section~\ref{sec:recommendations:apple-intelligence}).

As a recap, the bullet-point summary of our recommendations is as follows.\footnote{This bullet-point summary of our recommendations is identical to that in Section~\ref{sec:intro}.}

\subsection{Compatibility with E2EE Security}
\label{sec:recommendations:cake}   %

\begin{recommendation}[Training]
Using end-to-end encrypted content to train \emph{shared AI models} is not compatible with E2EE.

\end{recommendation}

As discussed in Section~\ref{sec:technical-implications:private-inference}, training a non-endpoint-local \emph{shared} model with E2EE data is incompatible with the security guarantees of E2EE encryption; regardless of how the data reaches the model or how the model is stored, the fact that other users can query a model that is a function of other users's E2EE data is not compatible with E2EE confidentiality. In practice, this can lead to leakage of sensitive data or malicious extraction via adversarial attacks.

It may be helpful to consider the following simpler model, based on an analogy. We can think of AI models as a ``bundle'' of the training data that went into them, which is somehow encoded in the values of its parameters---not necessarily perfectly encoded,\footnote{E.g., the bundle may be disorganized, and it may be difficult to extract every single piece of training data on demand.} but captured well enough that the model's outputs are highly dependent on this bundle of data. Then, an AI assistant trained on E2EE messaging data can be thought of as a \emph{bundle of private messages}. The model receives queries from users, performs a computation on this bundle of messages, and outputs a response. Through this lens, we can see that entities with access to the model itself---for example, the service provider, or if the model is stored directly in end-user devices---now have access to (a bundle of) messages not intended for them---a bundle of E2EE content---violating E2EE confidentiality. Further, even if no entity has direct access to this bundle (for example, if the model is somehow encrypted), any user can still \emph{perform computations} on this bundle of messages, and receive outputs that depend on it---which are E2EE content, being derivatives of E2EE plaintext messages---with just query access to the model. In other words, model outputs can be thought of as being ``tainted'' by any E2EE content in the underlying training data, and thus models trained on E2EE content produce outputs that are necessarily dependent on these training data and thus are themselves E2EE content.  So, training shared AI assistants with E2EE data \emph{definitionally} undermines the confidentiality guarantees of E2EE technologies.

\begin{recommendation}[Processing]
Processing \emph{E2EE content} for AI features (such as inference or training) may be compatible with end-to-end encryption only if the following recommendations are upheld:
        \begin{enumerate}
            \item Prioritize \emph{endpoint-local} processing where possible.
            \item If processing E2EE content for \emph{non-endpoint-local} models, 
                \begin{enumerate}[label=(\roman*)]
                    \item No \emph{third party} can see or use any \emph{E2EE content}\footnote{Or any derivative of E2EE content (i.e., any non-trivial function of content). See Section~\ref{sec:defs:e2ee-terms} for full definitions.} without breaking encryption,\footnote{This may sound paradoxical; see Section~\ref{sec:technical-implications:private-inference} for detailed discussion of technical configurations that may achieve this condition.} and
                    \item A user's \emph{E2EE content} is exclusively used to fulfill that user's requests.
                \end{enumerate}
        \end{enumerate}

\end{recommendation}

Application developers should prioritize local inference whenever possible, as this option is fully E2EE-compatible. %
To maintain E2EE confidentiality in non-endpoint-local inference, it is critical that no third party (including the model provider) handles any E2EE data unencrypted.\footnote{The one cryptographic tool that might permit encrypted processing, FHE, is not currently practical for this use case. See Section~\ref{sec:technical-implications:private-inference} for more detail.}

If exploring TEE-based approaches, service providers should not describe these as E2EE (see also Recommendation 3), and exercise great care around user expectations, consent, and unanticipated harms around transitioning a system from one type of security guarantee to another (see also Recommendation 4).\footnote{As discussed in Section~\ref{sec:background:trusted_hardware:tees_v_e2ee}, TEE security is different from encryption security: a system that relies on TEE security for message confidentiality does not achieve E2EE security and cannot accurately be described as offering E2EE.}

\subsection{Towards Meaningful Disclosure and Consent}
\label{sec:recommendations:discosure-consent}

In practice, legal consent and disclosure are often analyzed together. 
For the purposes of this section, we distinguish disclosure as the set of texts in advertising and user interfaces that the user is likely to read, despite formally not being part of a service's privacy policy or ToS. This allows us to recommend changes to the set of communications that may actually change some individuals' behavior. This kind of disclosure can also be described as one-way communication from the service to the user.

Consent, then, encompasses assent to the formal terms of service and privacy policies that bind the user and service in a contractual agreement. Opt-in consent mechanisms are those which allow users to take an affirmative step to agree to legal terms, such as pop-up banners for clicking a button or checking a box. These mechanisms may display brief disclosures informing users of the most important terms and may link to the pages where the complete terms and policies are displayed. Conversely, opt-out mechanisms are interfaces that provide consumers with a streamlined process to opt out of certain data processing activities after consent has been given, or as an alternative to the initial opting-in. Opt-out can therefore be considered withdrawal of consent. While users may not read all the fine print (see Section~\ref{sec:socio:what-users-understand}), the content contained in terms of service meaningfully affects users' rights.%
\footnote{Using these definitions, the line between disclosure and consent can be blurry in some situations. We include terms of service and privacy policies under the scope of consent and not disclosure. Elements of a service's chat interface, like pop-up notifications, are included in disclosure, but not necessarily consent. We treat advertising, which is generally not considered binding on a service, as a disclosure. Cookie banners, ubiquitous since the passage of GDPR, overlap both disclosure and consent: although easily dismissed by many consumers, they are prominently displayed on websites and contain non-legal language designed to be legible, a common characteristic of disclosures. Consumers then select their cookie preferences, making a two-way communication characteristic of consent.}

Next, we present our Recommendations 3 and 4, which are respectively focused on disclosure and consent.

\begin{recommendation}[Disclosure]\label{rec:disclosure}
Messaging providers should not make unqualified representations that they provide E2EE if the default for any conversation is that \emph{plaintext-dependent} versions of \emph{E2EE content} is used (e.g., for AI inference or training) by any \emph{third party}.
\end{recommendation}

``Unqualified representation'' is a term often used in consumer protection law. It generally refers to a statement made without caveats, and, in this paper, excludes caveats made in fine print, such as footnotes or terms of service. So, a splash page for an E2EE messaging service which boldly states ``E2EE Used for All Messages*'' would be making an unqualified representation, even if the asterisk points to a footnote that says ``Unless opted out by user.'' or ``Except when AI features are used.''

Put more simply, Recommendation 3 says that messaging services should not make unqualified claims that they are using E2EE when their implementation does not achieve E2EE confidentiality.

Moreover, as discussed in Section~\ref{sec:sociotechnical-implications}, to avoid regulator scrutiny for deceptive practices, 
messaging services hoping to make AI available to or train AI in E2EE contexts should err on the side of over-disclosure.

The requirements are clear for many possible implementations. A service with a configuration which falls into Category 1 (as defined in Section~\ref{sec:buckets}) would be able to advertise itself as using E2EE with no qualifications or caveats. 
On the other hand, services with configurations which fall into Categories 3–5 would need to make \emph{qualified} claims regarding their use of E2EE. %
Most Category 2 services will require additional, specific disclosures based on their features. 

\begin{figure}
    \centering
    \includegraphics[height=3in]{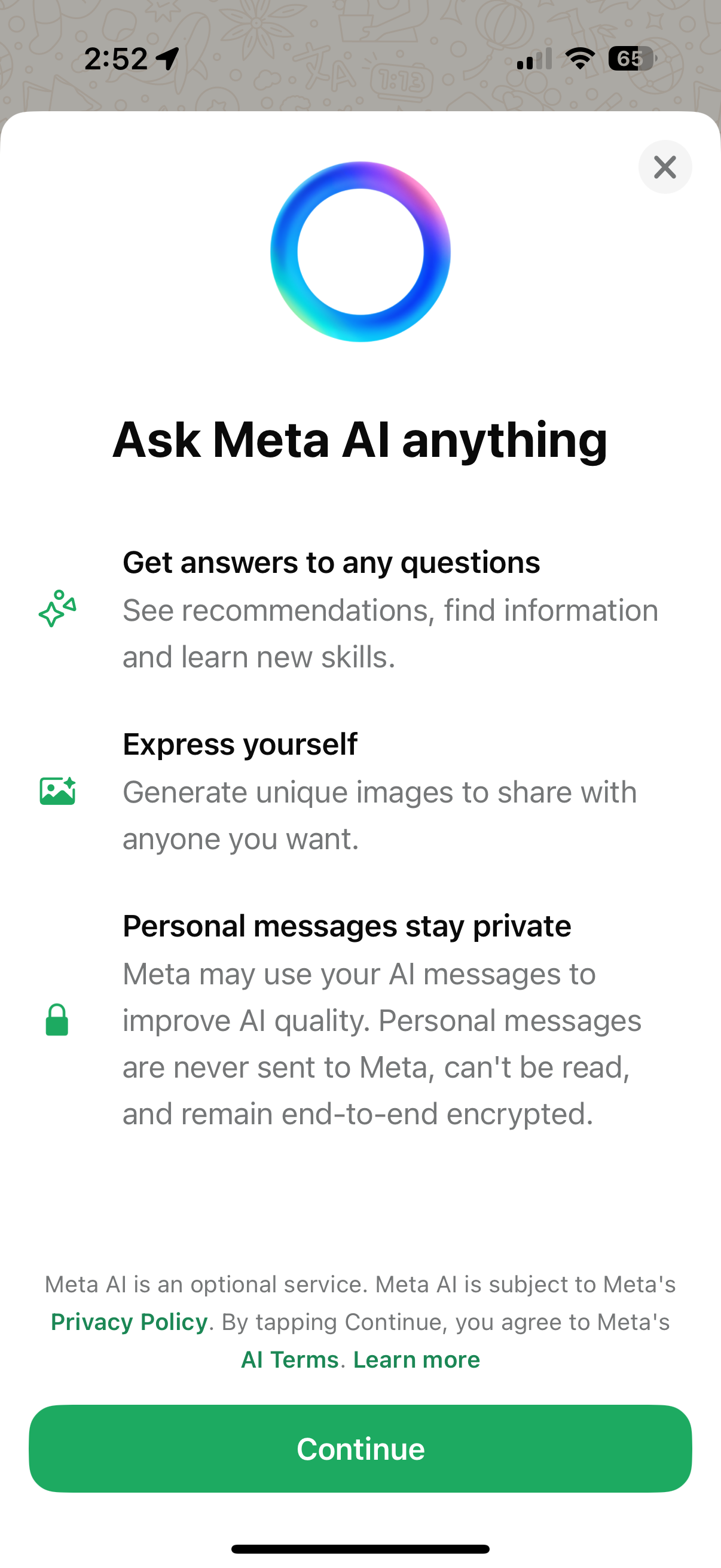}
    \caption{WhatsApp's disclosure upon a user's first use of Meta AI as of 15 November 2024}
    \label{fig:whatsapp-disclosure}
\end{figure}

For example, in WhatsApp, many points of the advertising and app interface indicate that ``text and voice messages'' are E2EE. Many users may see messages invoking Meta AI, which are not encrypted, as a ``text or voice message,'' and thus expect that their requests to Meta AI \emph{are} E2EE. Meta reinforces this false, but natural, misunderstanding. As shown in Figure~\ref{fig:whatsapp-disclosure}, when entering a Meta AI chat for the first time, WhatsApp users are shown a screen which states ``Personal messages stay private. Meta may use your AI messages to improve AI quality. Personal messages are never sent to Meta, can't be read, and remain end-to-end encrypted.'' 
The scope of ``personal messages'' is critical, and ambiguous, in this disclosure.
A sufficient disclosure would explicitly state that the messages in the following conversation are not encrypted and are sent to Meta, and as such should not include sensitive information.

Besides clarity and accuracy, another important component of disclosure is giving users notice of said disclosures. In other words, services should ensure that their disclosures are actually reaching users. Since E2EE services often do not require an email address to sign up, relevant disclosure updates should be sent to all users via the service itself.

\begin{recommendation}[Consent, opt-in and opt-out]\label{rec:opt-in-consent}
AI assistant features, if offered in E2EE systems, should generally be off by default and only activated via opt-in consent. Obtaining meaningful consent is complex, and requires careful consideration including but not limited to: scope and granularity of opt-in/out, ease and clarity of opt-in/out, group consent, and management of consent over time.
\end{recommendation}

Despite normative critiques highlighting the insufficiency of consent (see Section~\ref{sec:normativecrit}), consent is still the main legal tool currently available for giving users greater control over their privacy rights. Implementing consent interfaces well can make an important difference.

We recommend that opt-in consent be the standard mechanism for allowing messaging services to train AI on user data. Consent for using messaging services must not be conflated with consent for additional features such as AI assistants. Consent for separate features should be collected in separate instances, each with a specifically tailored opt-in banner or other form of disclosure. Although the banner should refer to a more detailed privacy policy and/or terms of service, there should be enough information on the banner to meet previously mentioned disclosure requirements. 

Messaging services should recognize the inherently private nature of communications, even (or especially) in environments that mix E2EE and non-E2EE data. Opt-in consent banners should serve the dual purpose of (i) clearly informing users about the exact scope of data that will be used for training purposes, including whether the data is protected by E2EE or not (or by some other security mechanism such as a TEE); (ii) providing users with a real choice, making sure to disclose how their choices may affect E2EE protection. It should be made clear to users which data is kept under E2EE (or TEE) protection, including any differences between individual and group settings. Additionally, terms of service and privacy or security policies should accurately and adequately describe security-relevant implementation details and associated risks of enabling the AI assistant feature. 

Beyond legal considerations, opt-in consent for AI features better preserves the application's systemic E2EE security properties (Section~\ref{sec:background:e2ee:systemic-e2ee-properties}): user inaction results in the unaltered functioning of E2EE protections, while only users who explicitly want the E2EE-violating feature will choose to have it enabled.

Finally, digital services should also provide clear opt-out mechanisms upon signing up and when new features are added, including a clear disclosure of what features still work if a user opts out. Each option should carry its own opt-out mechanism. A clear opt-out mechanism is one that is not concealed by dark patterns or unnecessary obstructions.\footnote{For instance, requiring users to navigate an eight-step process to access the opt-out form is not a clear mechanism \cite{fpfbrmeta}.} As a rule of thumb, opting out or withdrawing consent for services to which users have previously accepted must be made as easily accessible as the initial consent for using the feature. When possible, opting out from allowing AI training on one's data should not be a precondition to the use of an entire app or to the use of AI features in the app. 
Relatedly, services should consider separating the opt-in/opt-out for use of an AI feature and for training based on such use. Especially regarding AI training, it must be clearly disclosed that the opt-out is forward-looking: data which the company has already used for training will remain in their models, regardless of any opt-out exercised.\footnote{Implementing retroactive withdrawal is currently practically infeasible \cite{goel2024corrective}.}
Finally, consent should be subject to periodic renewals and opt-in banners should be re-validated after reasonable periods of time.

\subsection{Applying Our Recommendations to Apple Intelligence}
\label{sec:recommendations:apple-intelligence}
Having established our recommendations, we return to discuss the example of Apple (introduced in Section~\ref{sec:apple}).
Analyzing the strengths and shortcomings of Apple's system, though based on the limited information available, provides a helpful illustration of our recommendations in practice. 

Many parts of Apple's system are compliant with our recommendations, and provide a positive illustration of what can be done to preserve E2EE security when deploying AI assistants. Some parts of Apple's system raise significant concerns related to our recommendations, and other parts are not well enough documented to support a detailed analysis of compliance with our recommendations. 

Apple's system has three modes of processing, as described in Section~\ref{sec:apple}.
We proceed case by case.

\paragraph{Case I (Local Models)} In Case I, no user data is used for training~\cite{appledeviceandservermodels}, which is compliant with Recommendation 1. Further, all processing of queries---including, importantly, queries involving E2EE content---occurs locally on device \cite{applepressrelease}. Therefore, Case I is also fully compliant with E2EE security according to Recommendation 2, and features relying on Case I may be publicized as E2EE without qualifications according to Recommendation 3. Apple claims to prioritize Case I; ``many of the models'' run fully on-device and it is the ``more complex requests'' that require using Cases II or III~\cite{applepressrelease}---though it is not clear exactly what this means, the statement appears to align with Recommendation 2(a). Lastly, during its first release in iOS 18.1, Apple Intelligence was off by default, and had to be manually activated by users in their device settings~\cite{applereleasesupport}, which is in line with Recommendation 4. As of iOS 18.3, Apple Intelligence is on by default and ``integrated across features in your apps''~\cite{apple-intelligence-ios183}. This is in violation of our Recommendation 4.

\paragraph{Case II (Server Model)} Just like Case I, no user data is used to train Apple's PCC models~\cite{appledeviceandservermodels}, which is compliant with Recommendation 1. However, 
notwithstanding the security-conscious design choices within Apple's PCC, 
processing inference queries inside trusted hardware provides a different kind of security from E2EE security, and thus is not compatible with E2EE under Recommendation 2. Therefore, in line with Recommendation 3, Apple should not advertise features relying on Case II as E2EE without qualifications; rather, Apple should clearly articulate the limitations introduced by the TEE processing, and explicitly state the contexts in which E2EE applications (both native and third-party) no longer meet E2EE guarantees. Such notices should be prominently displayed. Importantly, even users who have Apple Intelligence off should be aware of these limitations, due to the systemic E2EE properties mentioned in Section~\ref{sec:background:e2ee:systemic-e2ee-properties}. Finally, during its first release in iOS 18.1, Apple Intelligence was off by default, and had to be manually activated by users~\cite{applereleasesupport}, which is in line with Recommendation 4. However, as of iOS 18.3, Case II no longer adheres to Recommendation 4, as Apple Intelligence is on by default~\cite{apple-intelligence-ios183}.

\paragraph{Case III (Third-Party Model)}
On Recommendation 1, if users do not sign in to their ChatGPT accounts, then OpenAI does not use user queries to train its models~\cite{applesupport-chatgptext}---thus, Case III processing is compliant with Recommendation 1 for users who are not logged in. However, if a user \emph{is} logged into their OpenAI account and has not configured their settings otherwise\footnote{ChatGPT accounts have model training on by default~\cite{openai-datacontrols}.}, queries outsourced to ChatGPT may be used for training OpenAI models~\cite{openaisupport-appleintelligence-datahandling}, directly opposing Recommendation 1. Thus, overall, Case III does not uphold Recommendation 1.

On Recommendation 2, if Siri is integrated within an application that is E2EE, then processing queries with ChatGPT via Siri would be in violation of Recommendation 2(b)(i), as E2EE content would be used by a third party. This is also true if ChatGPT can respond to queries involving third-party E2EE or ADP-E2EE application content. Furthermore, if a user has logged into their OpenAI account and has not configured their settings otherwise, ChatGPT can log requests and use query data to train its models. Using ChatGPT to process queries on E2EE content would therefore be in violation of Recommendation 2(b)(ii), since the user's E2EE content would not solely be used for the purpose of fulfilling that user's requests. 

Hence, according to Recommendation 3, Apple should not advertise features relying on Case III as E2EE without qualifications; it should clearly disclose the contexts in which E2EE data could be processed by third parties and how those third parties could use it.

Finally, since the ChatGPT extension in Siri is off by default and must be explicitly enabled in settings~\cite{applesupport-chatgptext}, Case III abides by Recommendation 4. Furthermore, it seems users are always made aware of when ChatGPT is being used to respond to a query~\cite{applesupport-siri}. Even if a user chooses not to be asked for permission before each query is sent to OpenAI, a small banner remains at the bottom of the response stating that it has been generated via ChatGPT.\footnote{For an example of what this looks like, see \href{https://www.apple.com/newsroom/images/2024/06/introducing-apple-intelligence-for-iphone-ipad-and-mac/article/Apple-WWDC24-Apple-Intelligence-Siri-ChatGPT-demo-240610_inline.jpg.large_2x.jpg}{this} example from the Apple Intelligence press release~\cite{applepressrelease}.} An additional disclosure that content sent to ChatGPT is no longer E2EE could further clarify security guarantees to users.

\section{Discussion}
\label{sec:discussion}

\subsection{Crypto Wars 3.0}
\label{sec:discussion:crypto-wars}

Law enforcement agencies and the global intelligence community have sought over decades to both limit the use of strong encryption and exploit the naturally occurring weaknesses of encrypted systems~\footnote{Very recent developments suggest an interesting possible shift, though it is too soon to be clear~\cite{CISA-FBI-2024,knodel-tpp24}. See also~\cite{baker19}.}~\cite{bellovin2013wiretapping}.
These efforts are in response to the technical reality that end-to-end encryption provides confidentiality to anyone who uses it, including criminals.
The so-called ``Crypto Wars'' can be described as the confrontation across regulatory, corporate, and research settings between state power and encryption proponents to determine whether there exists a technical compromise that would facilitate ``lawful access'' to confidential data~\cite{NAP25010}.
The Crypto Wars have evolved over the years with encryption, and technical proposals have evolved from pushing for weak cryptographic standards to various forms of key escrow~\cite{abelson2015doormats}.
The latest proposed technological solutions are framed as sophisticated alternatives to backdoors, which would preserve strong encryption algorithms while accessing E2EE content in other ways, namely on the device (the ``client'') and in the cloud using cryptographic computational techniques not requiring decryption.

What do backdoors, client-side scanning and homomorphic encryption have to do with AI assistants? Like data-driven algorithms, lawful access is incompatible with the privacy and confidentiality of end-to-end encryption, especially when users are uninformed of the risks. Furthermore, both are extensions or features of private messaging applications, not meant to enhance privacy and confidentiality; rather, they are designed to undermine it by enabling access to data that would not otherwise be accessible. A recurring concern in the Crypto Wars has been the risk of large-scale, hard-to-detect abuse and compromise: Even if we all agreed that AI assistants are beneficial, if E2EE data can be accessed by AI assistants for good, that method of access could also be used to track and target users, undermining the core of the security guarantee that E2EE provides.

\subsection{Business Incentives to Integrate AI (and Fundamental Rights)}
\label{sec:discussion:integrated-assistants}

Companies with data-driven business models may see messaging data as a valuable resource for refining AI algorithms, and the temptation to use this data will likely grow as AI assistants become increasingly embedded in everyday devices.
Messaging data is a source of largely untapped human-authored text at a time when concerns are mounting about the scarcity of such data; providing E2EE hinders the platform's ability to collect and analyze this data.

We are concerned that this new business incentive may eclipse the existing business incentives to protect user data, which are better aligned aligned with human rights protections. In what appears to be the new business model for Apple---which for years built its brand around privacy---and its competitors, services are seeking out and leveraging large sources of user data at a pace that does not permit thorough community deliberation over associated risks and externalities, even in the absence of clear user demand, and in the face of user concerns~\cite{cicek2024adverse,kikuchi2023growing}.

\subsection{But Can't I Just Paste That Into ChatGPT?}

Of course, a user can always take their E2EE messages and paste them into an AI assistant. That doesn't violate E2EE confidentiality. So why are we making such a big deal about the idea of E2EE content being sent from a user's device, where the user has them decrypted anyway, to an integrated AI assistant? 

At a glance, integrating AI features in an E2EE messaging app might seem like the same thing as pasting one's own messages into an AI assistant---just streamlined, which might be a convenient benefit for the user. In fact, we believe it is different in many ways, though a full exposition is beyond this paper's scope; interested readers may refer to the theories of privacy briefly discussed in Section~\ref{sec:sociotechnical:social-dimensions} for additional context.

In a nutshell, the two key differences are in the \emph{service provider's promise} about an E2EE application---that it does not access or process E2EE content, that it designs its systems to uphold E2EE confidentiality---and in the \emph{systemic nature of the breach of confidentiality}. If a user pastes their E2EE messages into an AI assistant, or indeed if they post them publicly online, that is a intentional act undertaken by an individual that may be a breach of social norms (depending on the context) but neither breaches the E2EE \emph{service provider's promise} or creates a \emph{systemic breach} of confidentiality by which similar processes can happen as a matter of course, as part of a regular app feature that is easily accessible and whose frequent use is encouraged by the very service provider who offers the E2EE.

\subsection{Additional Security Concerns: Message Integrity}
This work primarily focuses on the interplay between AI assistants and the confidentiality and systemic guarantees of E2EE.
Yet, as discussed in Section~\ref{sec:background:e2ee:auxiliary-e2ee-properties}, modern E2EE applications aim to provide additional security properties beyond these two.
An important direction for future work is to perform an analysis of the (in)compatibility of AI assistants and these other guarantees.
For any security property that an E2EE application claims to provide, it is critical that the implementation of AI assistants upholds this guarantee.

\emph{Integrity} is another standard security guarantee of E2EE applications, which states that a message received by the recipient should be equal to the one sent by the sender.
AI \emph{summary features} could be in tension with integrity: incorrect summaries of messages could be considered a violation of this guarantee, as they display an inaccurate representation of the message to the recipient.
While the original (unmodified) messages are still visible in the chat history, incorrect summaries may still cause unintended consequences akin to if the messages were modified in transit.
Incorrect summaries may occur accidentally, as a result of imperfect AI features, or deliberately, as a result of targeted attacks.
For example, consider a group chat with Alice, Bob, and Eve, where AI summaries are enabled.
If Alice is idle for a period of time while Bob and Eve send messages, and if Eve can trick the AI feature into misrepresenting Bob's messages in the summary, this could create confusion or conflict between Alice and Bob.

Whether and in what contexts incorrect AI summaries could or should be treated as a violation of integrity or not is an interesting question, outside the scope of this work.
The analysis is more subtle than for confidentiality, since integrity violations can in principle be detected by users by inspecting the original messages in the full chat history---in contrast, once E2EE content has been disclosed to third parties, it is not possible to backtrack.
Exploring this question may require a broader socio-technical treatment that analyzes normative expectations of integrity.
We anticipate that such tensions between integrity and AI assistants may become more relevant as AI summary features become more ubiquitous and more relied upon.

\section{Conclusion}
\label{sec:conclusion}

AI models are being developed rapidly and integrated across devices and applications, including in contexts where they may process end-to-end encrypted or otherwise private content.
In this work, we performed a detailed examination of complementary technical and legal tensions inherent to integrating AI models with E2EE applications, identifying some serious risks and concerns. 
Perhaps the greatest risk is in eroding E2EE as a systemic guarantee---a condition of any application or service that is held out to be E2EE---and turning it instead into merely a bonus feature that is routinely compromised in exchange for convenience. 

Development of AI features in E2EE environments should be accompanied by informed multidisciplinary discussions on potential harms. Our work represents a first step towards this; we hope this conversation continues, involving free speech, privacy, and human rights advocates, civil society representatives, and others outside academia, to develop frameworks and standards for the responsible deployment of AI and E2EE. It is critical for these debates to occur promptly, before unilateral industry rollout.

\section*{Author contributions}

MK had the initial idea and co-led the overall structuring of the paper and recommendations with SP. AF, BLH, and DY contributed equally to designing and writing technical background, implications, and recommendations. DF and JL contributed equally to designing and writing legal background, implications, and recommendations. SA led the case studies and provided research support throughout other aspects of the paper. SP was faculty lead and KC provided consultation.

\section*{Acknowledgments}

We are grateful to 
Peter Hall, 
Daji Landis, 
Julia Len,
Megan Richards, 
Nathalie Smuha,
and Michael A. Specter
for helpful discussion.

\printbibliography

\appendix
\stoptoc
\section{E2EE Features Extended Evaluation Framework}
\label{sec:e2ee-features-eval}

In the course of evaluating the many ways in which AI features might be introduced in E2EE applications, we created a framework that describes necessary considerations for determining whether a specific feature of an E2EE messaging application can be implemented such that it is still ``E2EE,'' which was helpful in guiding our analysis.
While we applied the framework largely to the context AI assistants in E2EE messaging applications, we believe there are many other features already widely deployed in mainstream E2EE applications whose implementations would benefit from use of this framework. AI has been a catalyst for our thinking about the role of data collection and use in today's communications apps.
While simple, we hope the framework facilitates a principled discussion of the implications of features in E2EE environments.

Our framework distinguishes between three core axes that must be considered in the context of application features: (1) the \emph{technical implementation} of the feature, and its confidentiality guarantees (or lack thereof). For considering AI assistants, we have performed this layer of analysis in Section \ref{sec:technical-implications:private-inference}; (2) the \emph{data inputs} to the feature, i.e., what type of user data will interface with it; and (3) the level of \emph{user control} over the feature.
We discuss these components in more detail below.
\betty{to confirm section placement before removing all mentions of AI assistants below}

\medskip
\paragraph{\textbf{Layer \#1: Technical implementation}} Does the underlying implementation of the AI assistant preserve the confidentiality guarantees of E2EE? %

We do so by presenting an abstract taxonomy of possible implementations, and labeling their compatibility with E2EE confidentiality. This captures the intuition that there are better and worse implementations of the same feature from the perspective of confidentiality, and that certain implementations do not admit any configuration that is confidentiality-respecting. For our purposes, there are two main confidentiality considerations for the implementation of an AI feature:
\begin{newitemize}
    \item \emph{Does the training of the AI system preserve the confidentiality of user data?} Training on personal data is fundamentally incompatible with E2EE, and thus no implementation of training on message data can guarantee confidentiality.
    \item \emph{Is inference performed in a way that preserves the confidentiality of user data?} Inference is in principle compatible with E2EE confidentiality, whether done locally (on-device) or through central servers.
\end{newitemize}

\medskip

\paragraph{\textbf{Layer \#2: Data inputs}} The next layer of our framework considers the data that is used as input to the AI features (for both training and inference). This layer is agnostic to a particular technical implementation, and is concerned exclusively with the data being used. There are four main considerations regarding the data that is fed to an AI features:
\begin{newitemize}
    \item \emph{What type of data is being used?} Types of data may include, for example, messages, links, images, user profile information, metadata, etc. Different AI features may be considered more or less sensitive due to the type of data it interacts with.
    \item \emph{Who is the intended recipient of the data?} The intended endpoint of communication may be either other users of the application, or the service provider itself. If it is the latter, then the data is not sensitive from the service provider's point of view. %
    \item \emph{What is the ``semantic'' value of the data?} The semantic value of data relates to the information content of the message. Data with higher semantic value is more sensitive. 
    \item \emph{Is the data referential?} Relevant to the prior point, some data may have ``intrisic'' meaning (e.g., messages), while other data may have ``referential'' meaning (e.g., URLs). Intrinsic data is generally more sensitive than referential data.
\end{newitemize}

\medskip

\paragraph{\textbf{Layer \#3: User control}} The last layer of our framework considers the level of control that the user has over an AI feature in regards to data privacy. There are four main considerations related to user control:
\begin{newitemize}
    \item \emph{Can the user opt out?} If the feature is on by default, does the user have the ability to turn the feature off? Being able to turn a feature off is more compatible with E2EE.
    \item \emph{What is the default configuration for the AI feature?} The AI feature can be either on or off by default within a user's application, with the latter being more compatible with E2EE than the former.
    \item \emph{What is the scope of the authorization?} Some features have \emph{full scope} of authorization, i.e., turning on the feature authorizes its usage across the entire application state. Conversely, features may have \emph{contextual scope} of authorization, in which case the feature must be turned on for individual application contexts. Contextual scope gives users more fine-grained control for how the AI assistant interacts with their data. For example, in a messaging application, it may be the case that the AI assistant must be individually authorized for each chat. Contextual scope is more compatible with E2EE.
    \item \emph{What is the barrier to toggling off?} Disabling an AI assistant can involve a range of friction and effort for the user, and require different levels of technical fluency. For example, an easy-to-find toggle button in the user settings requires less effort from the user and presents less friction than one that is buried in the settings, or that requires sending a request to the service provider. Low barriers to toggling off are more aligned with the principles of E2EE.
\end{newitemize}

These layers are roughly ordered from most technical and objective to the most context-dependent and normative. For example, the underlying implementation of a feature (Layer \#1) may reveal a user's data in plaintext, and therefore immediately preclude that feature from being compatible with E2EE just on the basis of the desired cryptographic guarantees. Conversely, arguing the acceptability of certain user control setups (Layer \#3) necessitates reference to normative principles.

\pagebreak
\section{Meta AI Disclaimers in WhatsApp}
\label{sec:metaai-disclaimers}

\begin{figure}[!htb]
\minipage{0.32\textwidth}
  \includegraphics[width=\linewidth]{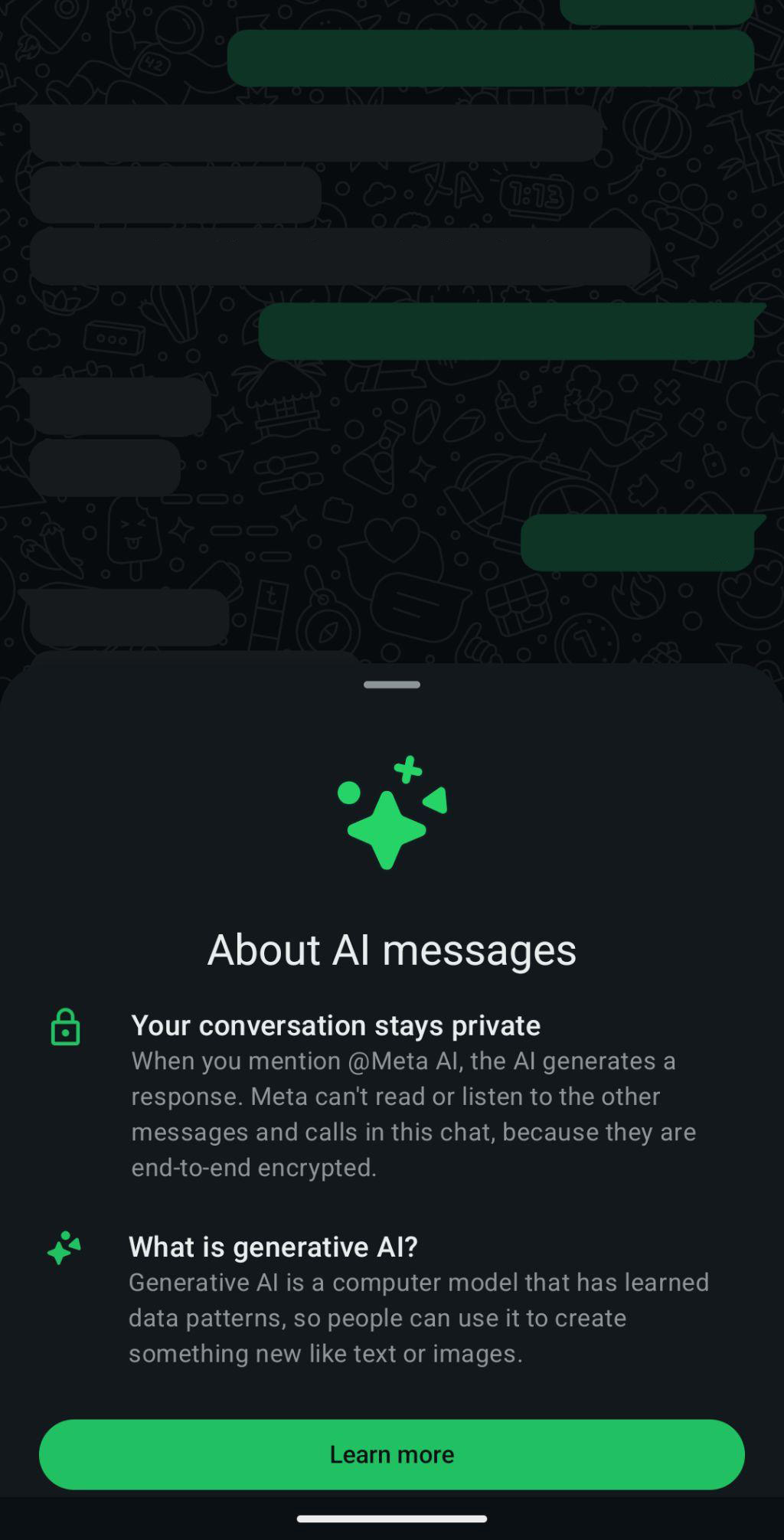}
  \caption{Meta AI's disclaimer when typing ``@MetaAI'' on WhatsApp for Android the first time, as of 02/14/25.}
\endminipage\hfill
\minipage{0.32\textwidth}
  \includegraphics[width=\linewidth]{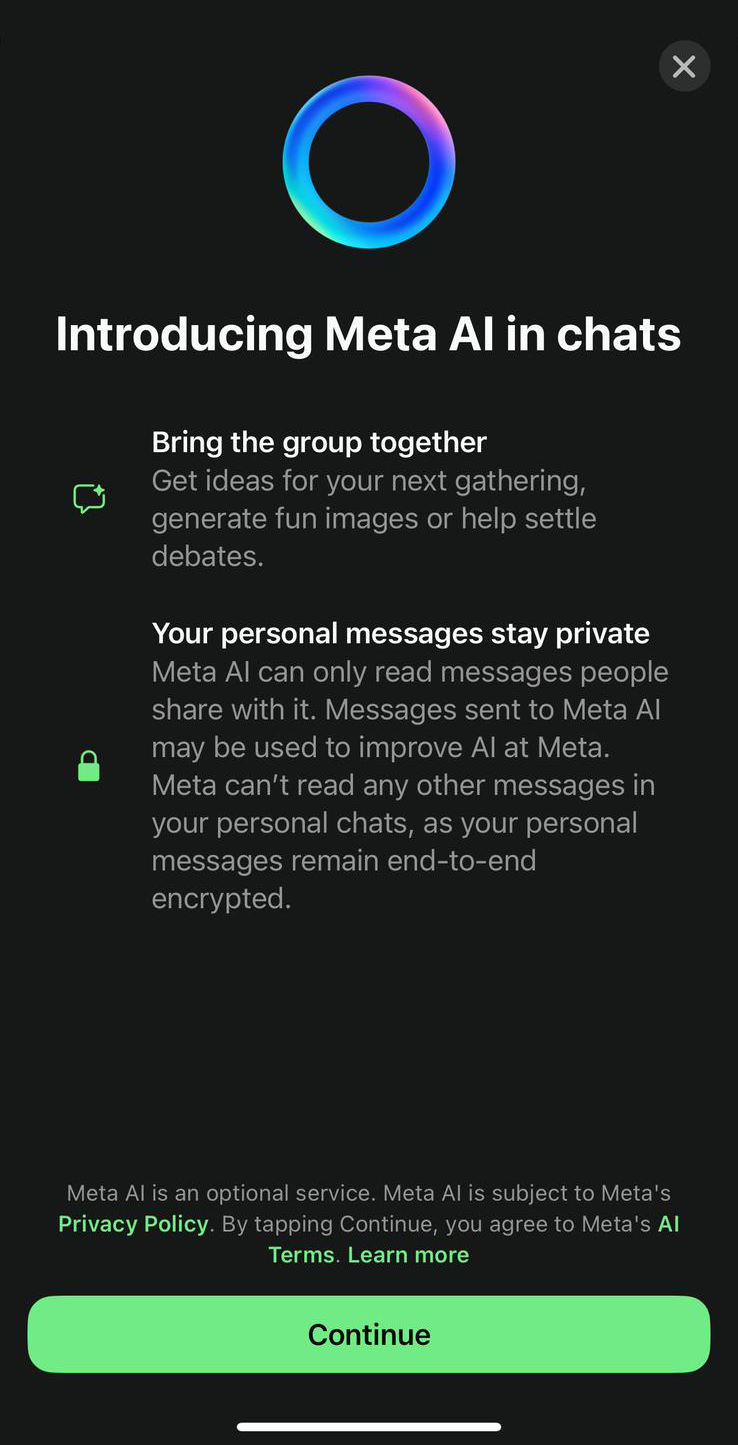}
  \caption{Meta AI's disclaimer when typing ``@MetaAI'' on WhatsApp for iOS for the first time, as of 03/15/25.}
\endminipage\hfill
\minipage{0.32\textwidth}%
  \includegraphics[width=\linewidth]{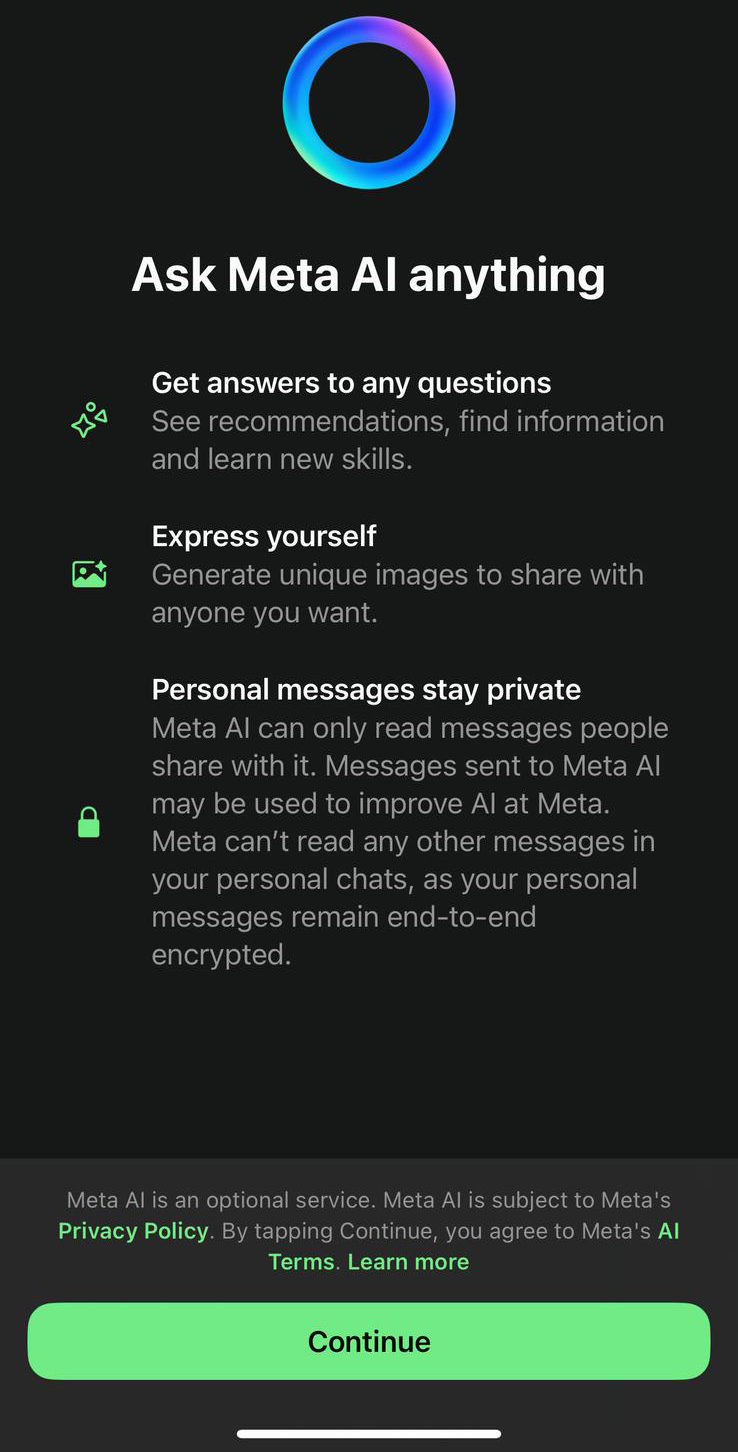}
  \caption{Meta AI's disclaimer when direct messaging with Meta AI on WhatsApp for iOS for the first time, as of 03/15/25.}
\endminipage
\end{figure}

\end{document}